\documentclass[usegraphicx,usenatbib]{mn2e}

\usepackage{times}

\usepackage[totalwidth=480pt,totalheight=680pt,paperwidth=210mm,paperheight=297mm]{geometry}

\usepackage{subfig}
\usepackage[countmax]{subfloat}
\usepackage{bm}% bold math
\usepackage{amssymb} % Defines special symbols such as ``\lesssim''
\usepackage[usenames]{color}
\usepackage[]{units}

\newcommand{\be}{\begin{equation}}
\newcommand{\ee}{\end{equation}}
\newcommand{\ud}{\mathrm{d}}

\title[Detecting the 21cm forest]
{Detecting the redshifted 21cm forest during reionization}

\author[Mack \& Wyithe]
{Katherine J. Mack$^1$\thanks{E-mail: mack@ast.cam.ac.uk}
and J. Stuart B. Wyithe$^2$\\
$^1$Kavli Institute for Cosmology, Cambridge / Institute of Astronomy, University of Cambridge, Madingley Road, Cambridge, CB3 0HA, United Kingdom\\
$^2$School of Physics, University of Melbourne, Parkville, Victoria, 3010, Australia
}

\date{Received 2011 October XX}

\pagerange{\pageref{firstpage}--\pageref{lastpage}} \pubyear{2011}

\begin{document}
\label{firstpage}
\maketitle

\begin{abstract}
The 21cm forest -- HI absorption features in the spectra of high-redshift radio sources -- can potentially provide a unique probe of the largely neutral intergalactic medium (IGM) during the epoch of reionization.  We present simulations of the 21cm forest due to the large scale structure of the reionization-era IGM, including a prescription for x-ray heating and the percolation of photoionization bubbles.  We show that, if detected with future instruments such as the Square Kilometer Array (SKA), the 21cm forest can provide a significant constraint on the thermal history of the IGM.  Detection will be aided by consideration of the sudden increase in signal variance at the onset of 21cm absorption. If radio foregrounds and the intrinsic source spectra are well understood, the flux decrement over wide bandwidths can also improve detection prospects. Our analysis accounts for the possibility of narrow absorption lines from intervening dense regions, but, unlike previous studies, our results do not depend on their properties. Assuming x-ray heating corresponding to a local stellar population, we estimate that a statistically significant detection of 21cm absorption could be made by SKA in less than a year of observing against a Cygnus A-type source at $z \sim 9$, as opposed to nearly a decade for a significant detection of the detailed forest features.  We discuss observational challenges due to uncertainties regarding the abundance of background sources and the strength of the 21cm absorption signal.
\end{abstract}


\section{Introduction}
\label{s:intro}

In both theoretical and observational astronomy today, a great deal of attention is being focused on improving our understanding of the epoch of reionization, when luminous sources began to ionize the intergalactic medium (IGM) for the first time after recombination.  Understanding the process of reionization would lend insight into not only the nature of the first sources themselves, but also the evolution of the IGM, the build-up of large-scale structures, and the radiative transfer processes occurring at high redshifts.  Currently, the most powerful data sets for constraining the ionization history of the IGM are the Gunn-Peterson absorption in the Lyman-$\alpha$ forest \citep{Bolton:2007} and the electron-scattering optical depth as observed in the cosmic microwave background (CMB) \citep{Larson:2010}.  The former places a lower limit on the fraction of hydrogen which is neutral (limited by the fact that the Gunn-Peterson trough saturates at a neutral fraction of $\sim 10^{-5}$) and the latter measures the integrated level of ionization between the observer and the CMB.  Thus we find a range of redshifts during which the universe went from substantially neutral to substantially ionized, but the detailed history of the reionization process is difficult to examine with these methods.

The hyperfine transition of ground-state neutral hydrogen (with a rest-frame wavelength of 21cm) has the potential to offer much more detailed information about the epoch of reionization than either the Lyman-$\alpha$ forest or the CMB optical depth measurement.  Since 21cm radiation is the result of a low-energy transition at radio wavelengths, it can free-stream to the observer without being substantially absorbed in the intervening IGM.  Also, most studies of the 21cm transition at high redshift focus on the signal in absorption or emission against the CMB, so the mean IGM itself rather than rare overdensities can be probed.

There has been a great deal of recent progress in observation and instrumentation aimed at the search for the redshifted 21cm signal from the epoch of reionization.  Several large-scale radio interferometers are currently being assembled to constrain the redshift of reionization as well as to produce all-sky brightness temperature measurements and a 21cm fluctuation power spectrum.  These instruments include the Low Frequency Array (LOFAR)\footnote{http://www.lofar.org/} in the Netherlands, the Giant Metrewave Radio Telescope (GMRT)\footnote{http://gmrt.ncra.tifr.res.in/} in India, the Experiment to Detect the Global EoR Signal (EDGES)\footnote{http://www.haystack.mit.edu/ast/arrays/Edges/}, the Precision Array to Probe the Epoch of Reionization (PAPER)\footnote{\citet{Parsons:2010}}, and the Murchison Widefield Array (MWA)\footnote{http://www.MWAtelescope.org/} in Western Australia.  Planning is also underway for the Square Kilometre Array (SKA)\footnote{http://www.skatelescope.org/}.

In this work, we discuss how studies of 21cm absorption against high-redshift radio sources can be used to study the evolution of the IGM around the epoch of reionization in what is known as the ``21cm forest'' \citep{Carilli:2002,Furlanetto:2002}.  In Section \ref{s:od}, we discuss how observations of the 21cm forest have potential advantages over other probes.  We describe in Section \ref{s:numsims} our simulation of the expected signal using an IGM evolution code.  In Section \ref{s:fx}, we discuss our parameterization of the x-ray heating of the IGM, and in Section \ref{s:sourcepop}, we discuss uncertainties regarding the population of high-redshift sources that could potentially be used as background sources for 21cm forest observations.  Using a hypothetical radio-loud high-redshift source as an example, we present sample 21m forest spectra in Section \ref{s:spectra}.  In Section \ref{s:stats}, we show how statistical detection methods can be used to isolate a faint signal in the case that radio-loud sources are rare at high redshift.  We present our conclusions and outlook for future work in Section \ref{s:conclusions}.

\section{21cm forest}
\label{s:od}

The optical depth to 21cm absorption of neutral hydrogen along the line of sight to a distant radio source depends primarily upon the density, temperature and ionization state of the gas.  For a cloud of gas at redshift $z$, with overdensity $\delta \equiv (\rho / \rho_0) - 1$ (where $\rho$ is the density and $\rho_0$ the mean density), neutral fraction $x_{HI}$, and spin temperature $T_S$, the optical depth to 21cm absorption is \citep{FOB,Xu:2009}:
\begin{eqnarray}
\tau_{\nu_0}(z) & = & \frac{3}{32 \pi} \frac{h_p c^3 A_{10}}{k_B \nu_0^2} \frac{x_{HI} n_H(z)}{T_S (1+z) (\ud v_\parallel / \ud r_\parallel)} \label{eq:FOB15} \\
 & \approx & 0.009 (1+\delta) (1+z)^{3/2} \frac{x_{HI}}{T_S} \nonumber \\
 &              & \times \left[ \frac{H(z)/(1+z)}{\ud v_\parallel / \ud r_\parallel} \right]. \label{eq:FOB16}
\end{eqnarray}
Here, $\nu_0$ is the observing frequency, $h_p$ is Planck's constant, $c$ is the speed of light, $A_{10}$ is the Einstein A coefficient, $k_B$ is the Boltzmann constant, $n_H$ is the number density of hydrogen atoms, $v_\parallel$ is the radial velocity of the gas, $r_\parallel$ is the distance to the gas and $H(z)$ is the Hubble parameter.  The factor in square brackets accounts for the peculiar velocity of the gas; for the purpose of this calculation, we will assume the gas is in the Hubble flow.  The spin temperature of the gas is defined by
\be \label{eq:spintemp}
T_S^{-1} = \frac{ T_\gamma^{-1} + x_\alpha T_c^{-1} + x_c T_K^{-1} }{ 1 + x_\alpha + x_c },
\ee
where $T_\gamma$ is the temperature of the CMB, $T_c$ is the color temperature of the Lyman-$\alpha$ background, $T_K$ is the gas kinetic temperature, and $x_c$ and $x_\alpha$ are coupling coefficients for collisions and the Wouthuysen-Field effect, respectively.

If a radio source is observed to exist in or before the epoch of reionization, the structure of the neutral hydrogen gas along the line of sight to the source should be apparent in 21cm absorption in the radio spectrum.  Several distinct features may be seen: a flux decrement blueward of the redshifted 21cm absorption frequency, whose depth depends upon the mean optical depth of the IGM at that redshift; small-scale variations in flux due to fluctuations in the density and ionization state of the IGM along the line of sight; transmission windows due to photoionized bubbles along the line of sight; and deep absorption features due to dense neutral hydrogen clouds in dwarf galaxies and minihalos.  The spectral features due to 21cm absorption are known, collectively, as the 21cm forest, in analogy to the Lyman-$\alpha$ forest.  In contrast to the Lyman-$\alpha$ forest, however, the 21cm forest occurs at significantly higher redshifts ($z \gtrsim 7$) and contains detailed information about the IGM even when the IGM is largely neutral. In the Lyman-$\alpha$ forest, the Gunn-Peterson trough reaches zero flux when the ionized fraction of the intervening gas reaches approximately 1 part in $10^5$.  Also, the 21cm forest has the potential to probe very small structures (limited primarily by the frequency resolution of the instrument) at lower overdensities than the Lyman-$\alpha$ forest.  Much of the information in the 21cm forest will come from structures that are still in the linear regime, potentially making it possible to probe the matter power spectrum more directly than would be possible via the Lyman-$\alpha$ forest.

As we discuss below, the utility of the 21cm forest as a diagnostic of the IGM depends upon both the evolutionary history of the IGM in heating and ionization and the abundance of appropriate background sources at high redshift.  We show that, should sufficiently bright background sources be found, observations of the 21cm forest with instruments such as the SKA can be useful in constraining the thermal history of the IGM as well as probing the growth of structure at high redshift.

\section{Numerical simulations}
\label{s:numsims}

To simulate the 21cm forest spectrum, we combine the results of two numerical simulations.  The first is a time-evolution code tracking the mean properties of a uniform IGM over cosmic time, and the second a non-uniform two-phase IGM simulation describing small-scale spatial inhomogeneities.  The mean IGM code is an adaptation of a code developed by \citet{MackWesley}.  It includes cosmological expansion, adiabatic cooling, spin temperature coupling with the CMB at high redshift, and x-ray heating of the gas via a star formation rate derived from calculations in \citet{Geil:2008}.  The relationship between the star-formation rate (SFR) and the x-ray luminosity is defined by \cite{Furlanetto:2006b}:
\be \label{eq:fx}
L_X = 3.4 \times 10^{40} f_X \left( \frac{\textrm{SFR} }{ 1 \textrm{ M}_\odot \textrm{ yr}^{-1} } \right) \textrm{ erg s}^{-1} ,
\ee
where $f_X$ is an x-ray efficiency factor.  We assume a spectral index of $\alpha=1.5$ in the frequency range between 0.2 keV and 10 keV for x-ray sources.  The factor $f_X$ is essentially a free parameter.  In this section, we assume $f_X=1$, but in subsequent sections we vary this quantity.  (See Section \ref{s:fx} for more details.)  The two-phase IGM code was developed by \citet{Geil:2008} and includes a semi-analytic prescription for star formation and subsequent photoionization.  The size of the simulation box is 70 cMpc cubed, with 360 pixels per sidelength.  We piece together four separate (arbitrarily chosen) 1D slices through the simulation box in order to include a range of local environments through which the line of sight may pass, giving a total of 280 cMpc.  Due to the nature of this study as a discussion of the feasibility of future measurements rather than a detailed quantitative analysis of simulation results, a dedicated, high-resolution simulation of the IGM was not necessary.

For the purpose of illustrating the basic properties of the 21cm forest for a source at high redshift, we first evolve the uniform IGM from high redshift to $z=7$ and calculate the optical depth to 21cm absorption for the range $7 \le z \le 25$.  Then, using the mean gas temperature and ionization fraction of the uniform IGM, combined with the density fluctuations and HII regions from slices through the non-uniform two-phase simulation, we produce a simulation of the small-scale variations in 21cm absorption in a restricted range from $z=8$ to $z=9$.  This allows us to produce 21cm forest spectra against a hypothetical $z=9$ radio source over a wide range of frequencies, with x-ray heating of the gas taken into account. Large spatial inhomogeneities in the gas temperature are unlikely to significantly alter our results, as the the mean free path of the x-rays responsible for the heating is generally comparable to or greater than the size of the non-uniform simulation box.

Note that our simulation includes large scale features of the IGM such as photoionized bubbles, but does not include collapsed structures such as galaxies and minihalos, which could be expected to produce deep absorption spikes in the 21cm forest spectrum \citep{Carilli:2002, Carilli:2004, Xu:2009, Xu:2011}.  There have been some estimates as to the number and size of absorbers that should be found in the IGM at these redshifts \citep[e.g., ][]{Furlanetto:2006a}, but these depend strongly on modelling not only small-scale overdensities and their gas properties, but also their heating and photoevaporation.  In this study, since our main interest is in the mean properties of the IGM, the presence of small-scale absorption features would complicate the mean-IGM signal, rather than being of interest as features themselves.  To account for the effect absorption lines would have in observations of a distant source spectrum, we have developed a procedure for stochastically inserting absorption lines into the spectra of our hypothetical sources.  This method is discussed in more detail in Section \ref{s:spectra}.

\begin{figure}
\centering
\subfloat[]
{\label{sf:xi}
\includegraphics[width=0.7\columnwidth,angle=-90]{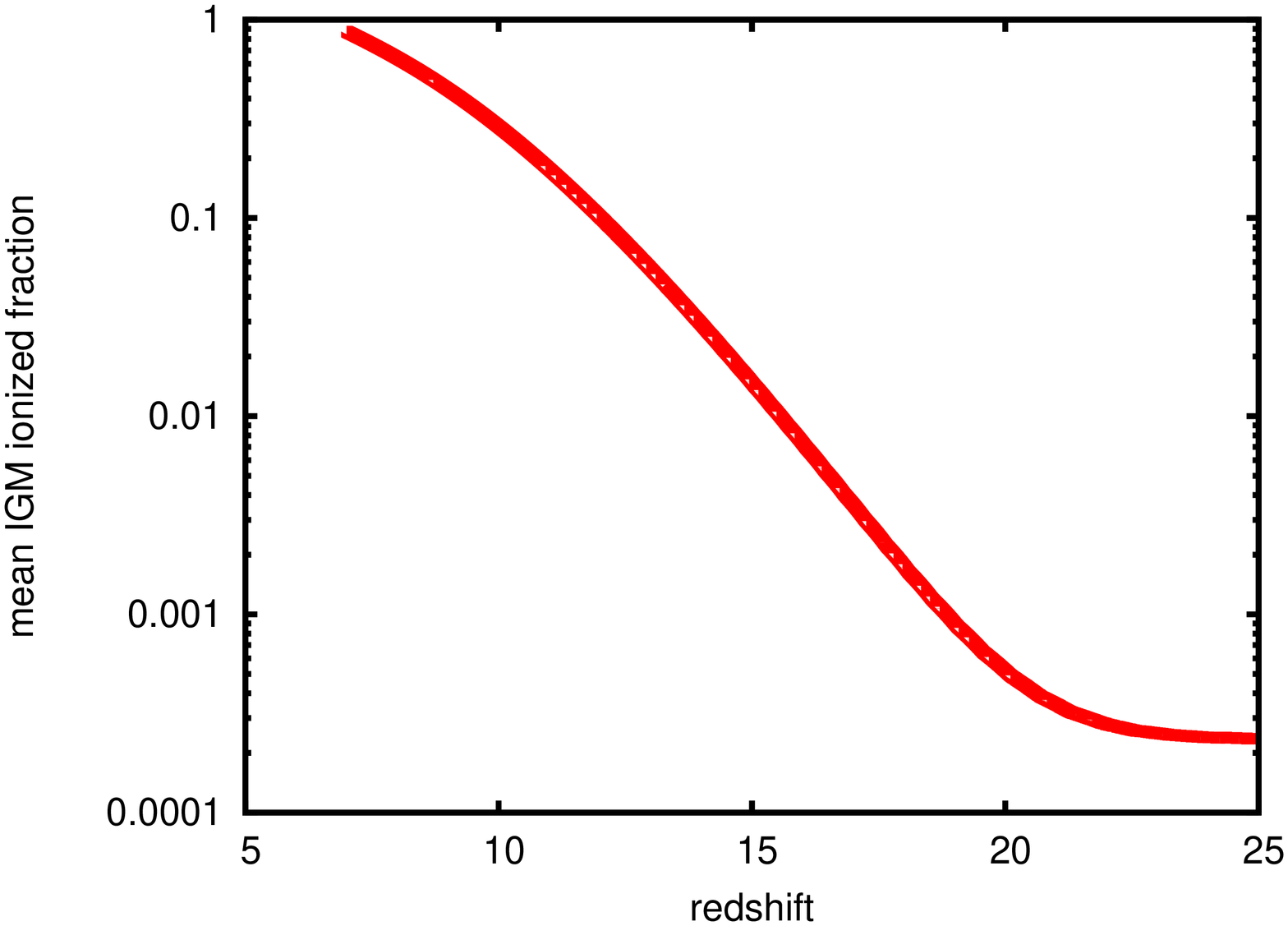}}

\subfloat[]{
\label{sf:TK}
\includegraphics[width=0.7\columnwidth,angle=-90]{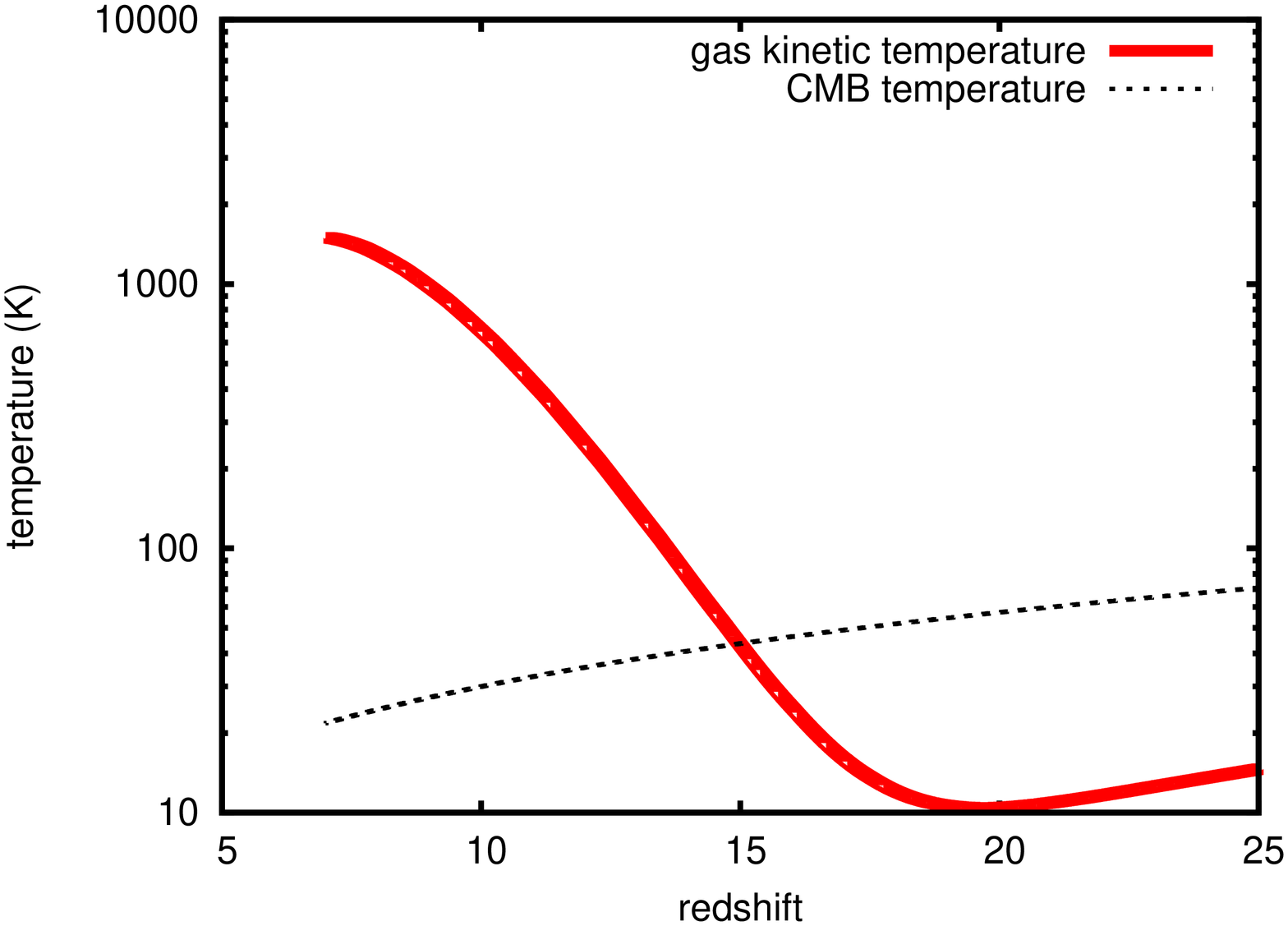}}

\subfloat[]{
\label{sf:meantau}
\includegraphics[width=0.7\columnwidth,angle=-90]{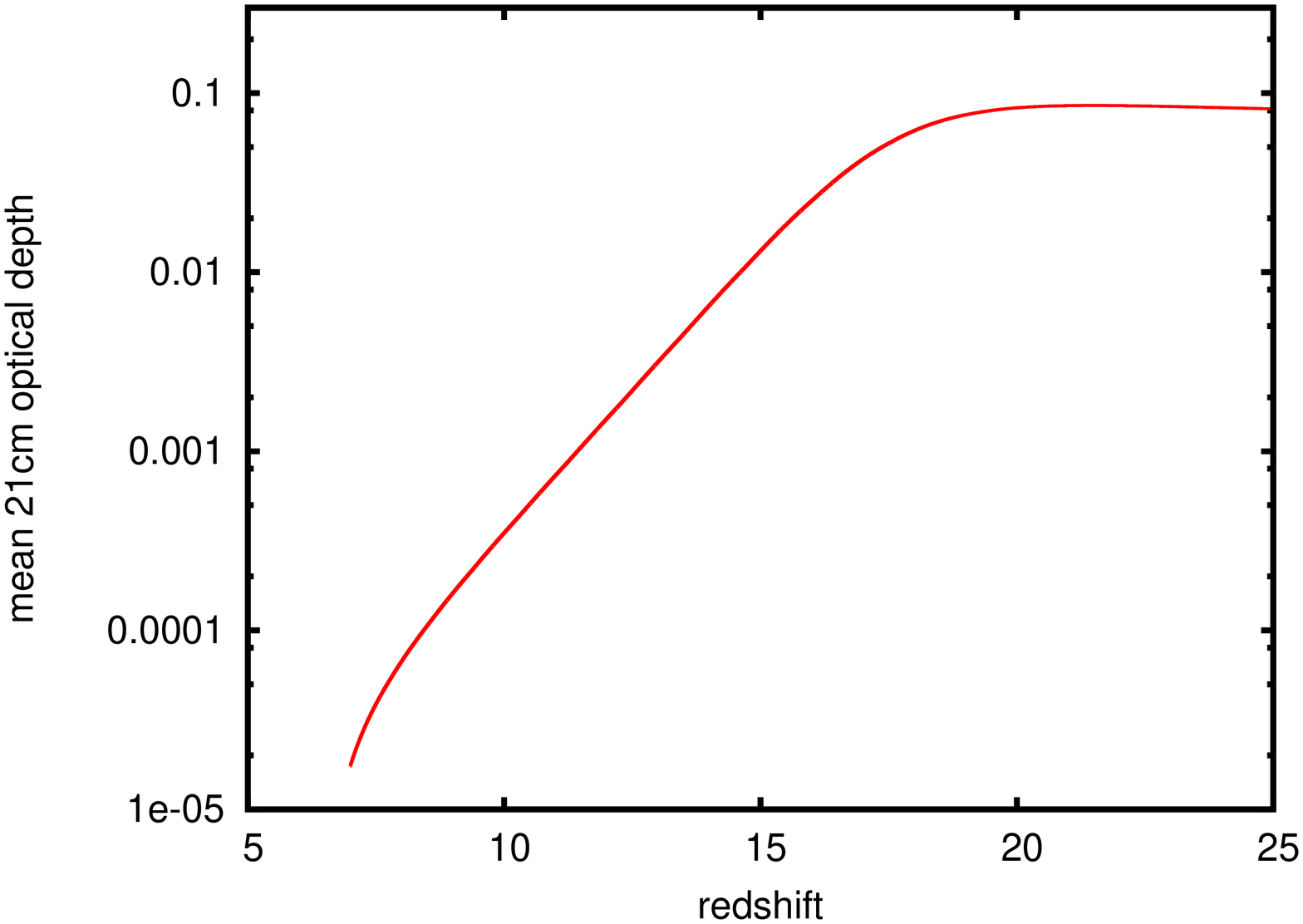}}

\centering
\caption{\label{f:uniformIGMsim}  Uniform IGM simulation output plots. \subref{sf:xi} Mean IGM ionized fraction $x_i$ as a function of redshift.  \subref{sf:TK} Gas kinetic temperature (solid line) and CMB temperature (dotted line) as a function of redshift.  \subref{sf:meantau} Mean 21cm optical depth as a function of redshift.  The model shown here has $f_X=1$ (see Section \ref{s:fx}).}
\end{figure}

\begin{figure}
\centering
\subfloat[]
{\label{sf:delta}
\includegraphics[width=0.7\columnwidth,angle=-90]{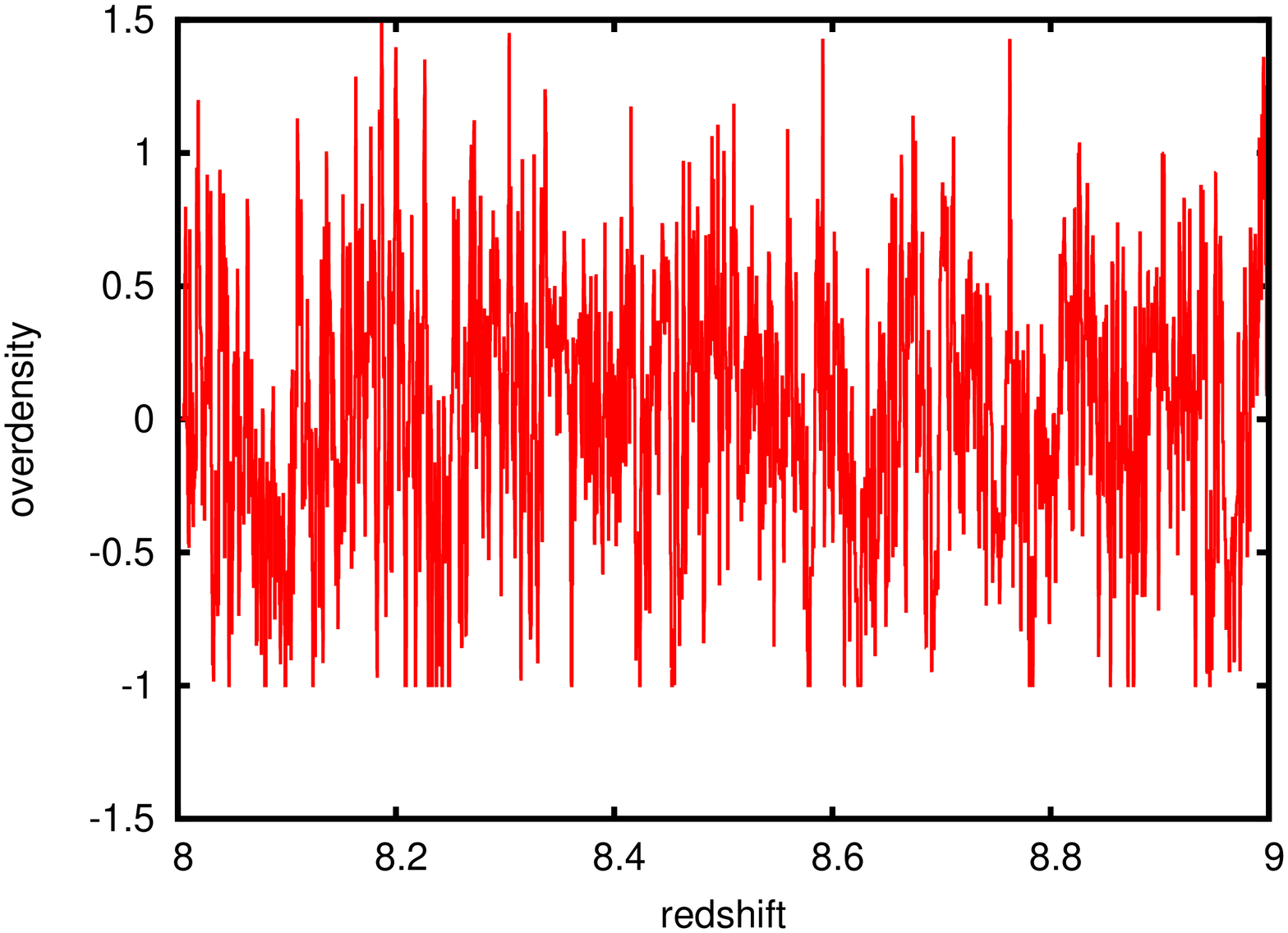}}

\subfloat[]{
\label{sf:xHI}
\includegraphics[width=0.7\columnwidth,angle=-90]{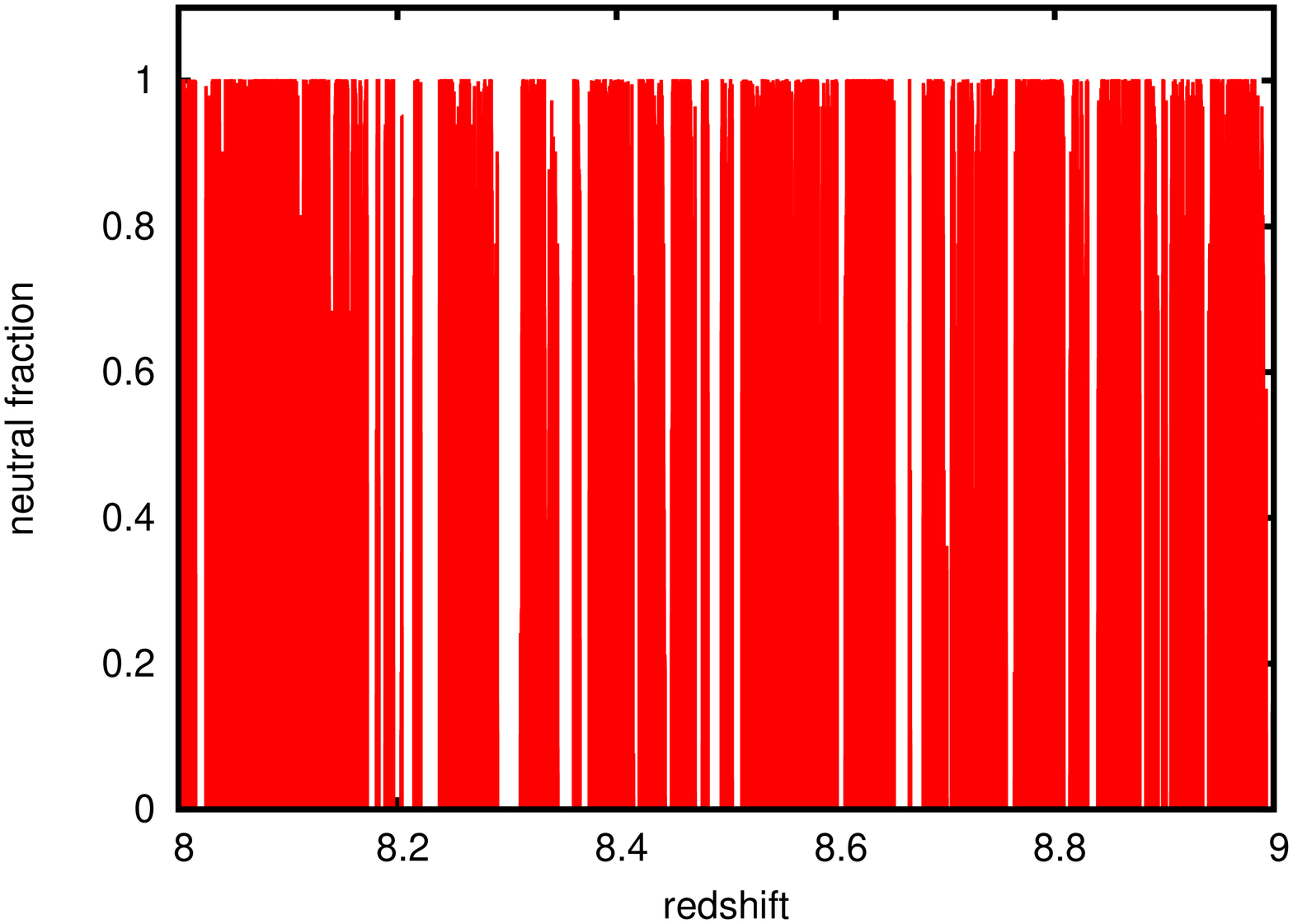}}

\centering
\caption{\label{f:hiresIGMsim}  Non-uniform IGM simulation output plots.  \subref{sf:delta} Overdensity $\delta$ and \subref{sf:xHI} neutral hydrogen fraction $x_{HI}$ in the non-uniform simulation region as a function of redshift, from a  two-phase IGM code developed by \citet{Geil:2008}.  The size of the simulation box is 70 cMpc cubed, with 360 pixels per sidelength. Multiple slices through the simulation box are used, giving a total of 280 cMpc.
}
\end{figure}

In Figures \ref{f:uniformIGMsim} and \ref{f:hiresIGMsim}, we plot selected quantities output by the IGM simulations.  Figure \ref{sf:xi} shows the mean ionized fraction in the uniform-IGM simulation.  The simulation has been calibrated for the overlap phase to occur at redshift $z=6$.  Reionization in this calculation is achieved primarily through the growth of photoionized regions due to star formation, although the small contribution from x-ray heating is also taken into account.  For the purpose of this study, we do not present alternative reionization histories or morphologies, but rather consider an example in which our calculations can be used to test the feasibility of future measurements.  It has been shown \citep{McQuinn:2007} that the neutral hydrogen fraction is the most important determinant of the structure of ionized regions during reionization; therefore, it should be straightforward to generalize our basic results by referring to the IGM ionization evolution we present in Figure \ref{sf:xi}.

In Figure \ref{sf:TK}, we plot the evolution of the kinetic temperature of the gas ($T_K$, solid line) along with the CMB temperature ($T_\gamma$, dotted line) in the uniform-IGM simulation.  The spin temperature of the gas ($T_S$) is coupled to the CMB at high redshift and then to the gas temperature by $z \lesssim 17$.  In our calculations of the optical depth, we assume $T_S = T_K$, which is a good approximation at the redshifts where a detection of the 21cm forest is feasible \citep{Santos:2008}.  In Figure \ref{sf:meantau}, we plot the mean 21cm optical depth as a function of redshift from the uniform-IGM simulation.  In this plot, the optical depth $\tau$ is somewhat overestimated for $z \gtrsim 17$ due to the $T_S=T_K$ approximation \citep{Santos:2008}.

In Figures \ref{sf:delta} and \ref{sf:xHI}, we plot the overdensity $\delta$ and the neutral fraction $x_{HI}$, respectively, in the non-uniform simulation region as a function of redshift.  The algorithm used to calculate overdensity allows for values $\delta < -1$; we have removed these unphysical overdensities by placing a cutoff at $\delta=-1$.  The plot of $x_{HI}$ is on a linear scale to emphasize an important feature of the non-uniform simulation: the presence of photoionized bubbles.  We have filled the region below the curve to show the contrast between regions of low ionization (which appear solid) and fully-ionized bubbles (which show up as white gaps).  These bubbles have a distinct signature in the 21cm absorption spectrum against a background source, which we will discuss in more detail in Section \ref{s:variance}.

%%%% 
%%%% 
%%%% 

%%%% 

\section{X-ray heating efficiency}
\label{s:fx}

The details of the heating of the IGM over cosmic history are still fairly unconstrained \citep{Mitra:2010, Pritchard:2008, Furlanetto:2006b}.  To account for possible differences in the thermal history of the IGM, we parametrize the heating by the x-ray efficiency factor $f_X$, defined in equation (\ref{eq:fx}). This is a free parameter, relating to the efficiency with which the x-rays produced by star formation couple to the IGM. We take $f_X$ to be anywhere from $0.01$ to 100, which is consistent with other recent works, e.g., \citet{Pritchard:2010}.  In the local universe, $f_X \approx 1$.   Although a direct link between $f_X$ and a specific physical model of star formation and radiative transfer is not straightforward, constraining $f_X$ (to within an order of magnitude, for instance) would be a helpful guide to future model building.

\begin{figure}
\centering
\subfloat[]
{\label{sf:mean21od}
\includegraphics[width=0.72\columnwidth,angle=-90]{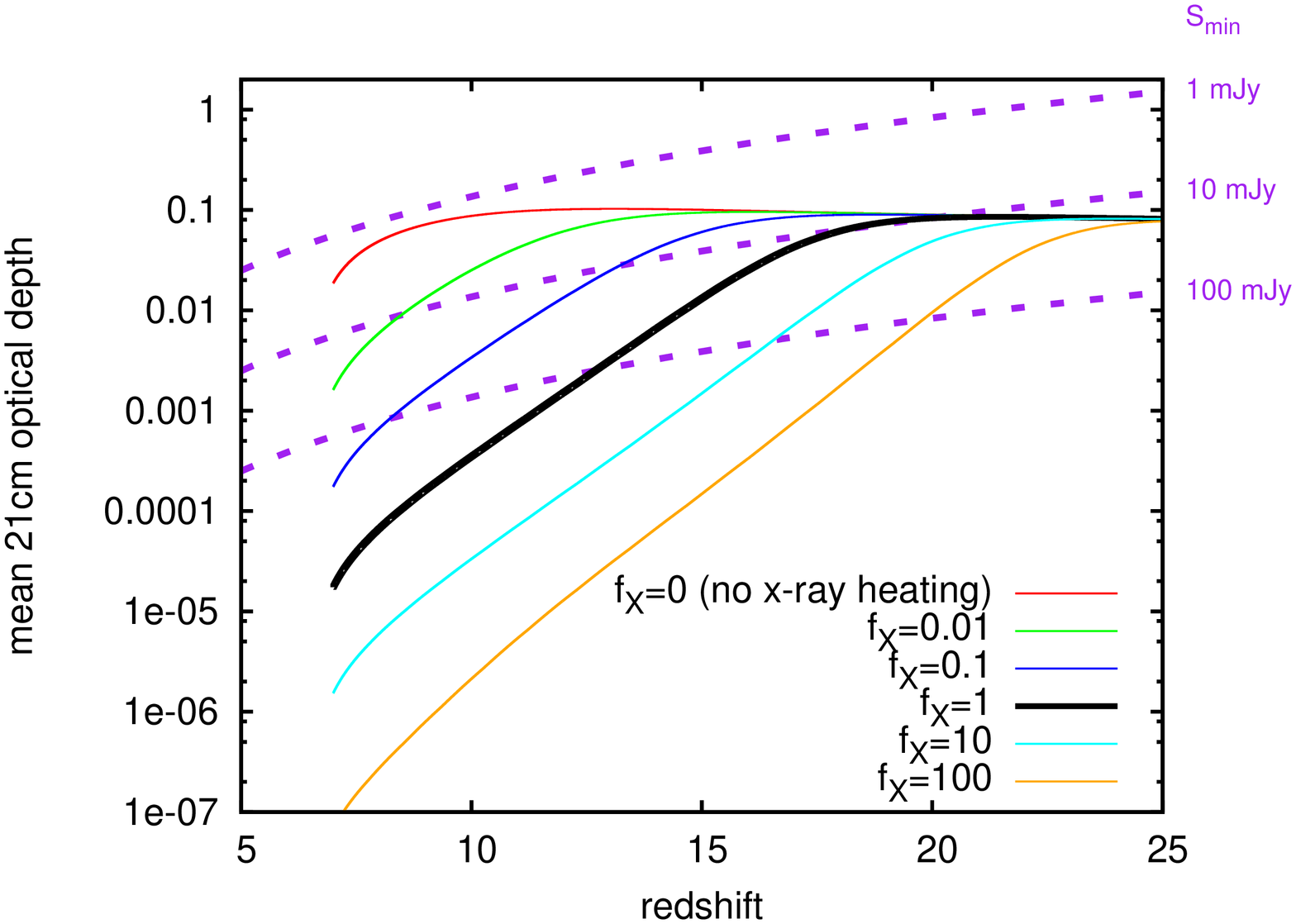}}

\subfloat[]{
\label{sf:fvaryingforest_freq}
\includegraphics[width=0.7\columnwidth,angle=-90]{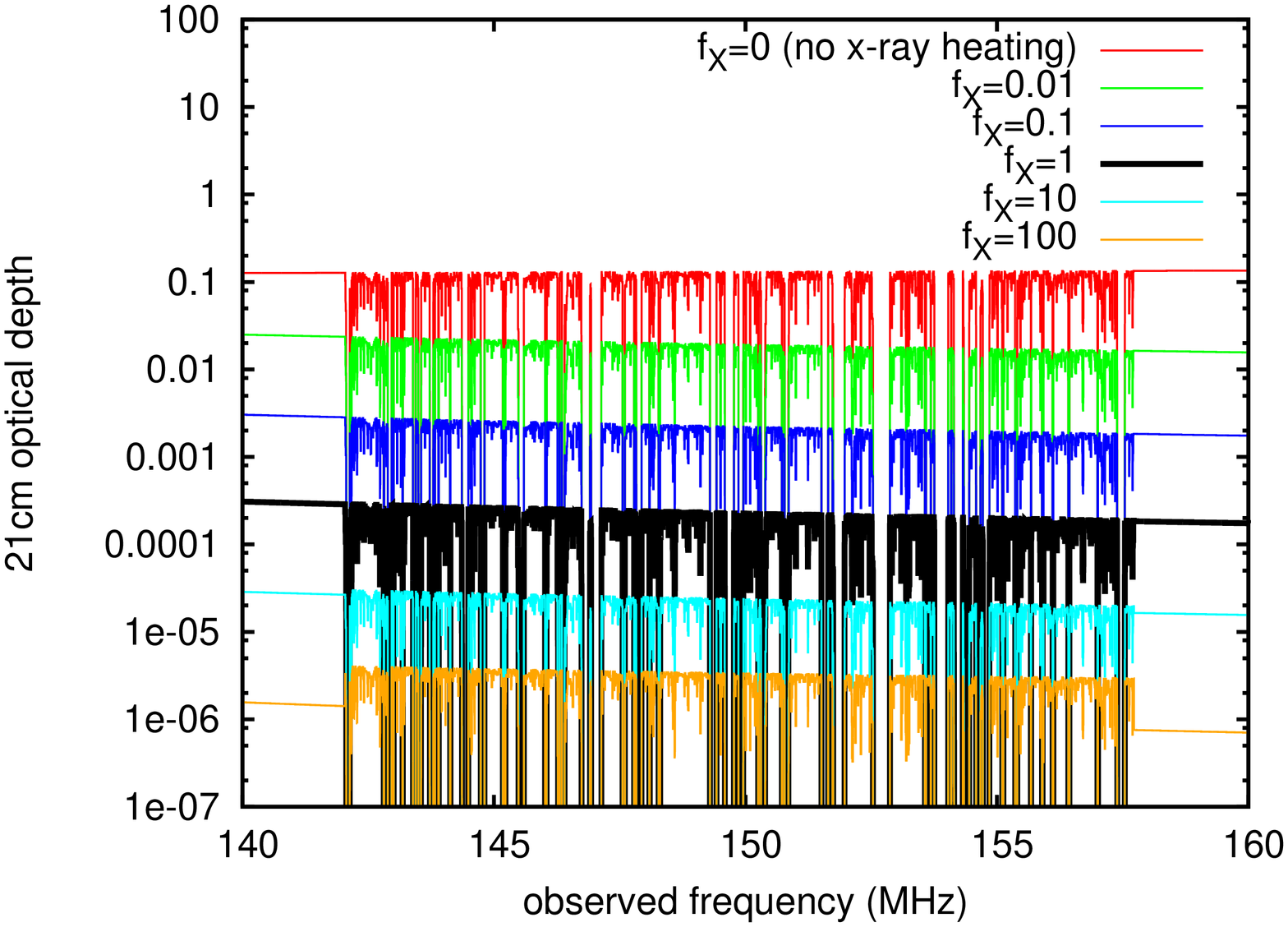}}

\centering
\caption{\label{f:opticaldepth}  21cm absorption optical depth.  \subref{sf:mean21od} Mean 21cm optical depth $\tau$ as a function of observed frequency, for varying $f_X$ values: from top to bottom curve, $f_X =$ 0 (no x-ray heating), 0.01, 0.1, 1 (thick black line), 10, 100.  Also included are curves of $S_{min}$ (dashed purple lines) as defined in equation (\ref{eq:Smin}).  These lines indicate, for each value of $\tau$ on the left-hand axis, the minimum observed flux density of a source that would allow a detection of absorption at signal-to-noise of 5, assuming an array with effective area $A_{eff}=10^6 \textrm{ m}^2$, frequency resolution $\Delta \nu_{ch} =$ 1 kHz, system temperature given by equation (\ref{eq:systemp}), and integration time $t_{int} =$ 1 week.  The redshift dependence comes from the frequency dependence of the system temperature.  Where the mean optical depth lines (solid) cross above the $S_{min}$ lines (dashed), the absorption is detectable at that redshift. \subref{sf:fvaryingforest_freq} 21cm forest optical depth as a function of redshift, for varying $f_X$ values: from top to bottom curve, $f_X =$ 0 (no x-ray heating), 0.01, 0.1, 1 (thick black line), 10, 100.}
\end{figure}

In our simulation, the main effect of the variation of $f_X$ is a change in the gas temperature, and, thus, the mean 21cm optical depth.  In Figure \ref{sf:mean21od}, we plot the mean 21cm optical depth as a function of redshift for $f_X$ values between 0.01 and 100.  The thick black line is for $f_X=1$, as in Figure \ref{sf:meantau}.  The lines with higher (lower) optical depth have lower (higher) $f_X$ values, representing less (more) effective heating.  We also plot, in Figure \ref{sf:fvaryingforest_freq}, the optical depth in the ``forest region'' where the non-uniform simulation determines the overdensity and ionized fraction.  In this figure, we assume a background source at $z=9$.  The low-frequency cut-off in the forest is due to the frequency of onset of 21cm absorption, and the high-frequency cut-off is due to the finite extent of the non-uniform simulation region.  From a close inspection of Figures \ref{f:hiresIGMsim} and \ref{sf:fvaryingforest_freq}, it can be seen that the variation in neutral fraction -- specifically the presence of photoionized bubbles -- dominates the spatial variation in the 21cm optical depth, whereas the density fluctuations are subdominant.  We also see from Figure \ref{f:opticaldepth} that varying $f_X$ by an order of magnitude can have a roughly order-of-magnitude effect on the optical depth. This is due to the fact that the overall optical depth is strongly dependent on the spin temperature $T_S$, which varies by roughly an order of magnitude with each factor of 10 in $f_X$.

From Figures \ref{sf:mean21od} and \ref{sf:fvaryingforest_freq}, we see that if the 21cm forest is detected in the spectra of high-redshift radio sources, measurements of the optical depth can constrain the parameter $f_X$ and thus give us insight into the thermal history of the IGM.  Against a sufficiently bright source, the 21cm forest spectrum can reveal the layout of photoionized bubbles along the line of sight (which would appear as transmission windows between regions of absorption), and, potentially, variations in density and ionized fraction.
Narrow absorption lines from dense neutral regions would also be detectable in such observations.  Absorption lines can have optical depths at levels as high as a percent \citep{Furlanetto:2006a}, but because of their narrowness (typically 2-3 kHz) their detectability in low mean optical depth regions would be highly dependent on their number and density.  In principle, $f_X$ can also affect the number and depth of absorption lines, by altering the background radiation field in which small halos or protogalaxies reside. We discuss the implications of absorption lines in Section 6.

For some redshifts and $f_X$ values, the mean optical depth is so low as to make detection of 21cm forest features difficult or impossible, even with future instruments such as SKA.  For a very radio-loud source at high redshift, a wide range of thermal histories will produce a detectable forest signal, and the optical depth can be used to constrain $f_X$, but sources with low flux density in the frequencies of interest may result in a detection only for low values of $f_X$.  In this case, a non-detection of the 21cm forest may provide a constraint on the thermal history.

In Section \ref{s:sourcepop} and \ref{s:spectra} below, we discuss how prospects for the detection of the 21cm forest depend upon the source population and the thermal history of the IGM, and we consider how the presence of narrow absorption lines can complicate the extraction of information about the thermal history.  Then, in Section \ref{s:variance}, we discuss methods for statistical detection in cases where the absorption spectrum has low signal-to-noise.

%%
%%%%
%%

\section{Source population and detectability}
\label{s:sourcepop}

The biggest challenge for doing observations of the 21cm forest will be finding sufficiently distant, radio-loud sources.  The optical depth to 21cm absorption increases with the redshift, density and neutral fraction of the intervening gas, but decreases with the gas's spin temperature.  Therefore the strongest signals will be found in the spectra of very high redshift sources -- those occurring prior to the epoch of reionization.  However, the detection of 21cm absorption also becomes more difficult as the radio flux density of the source decreases: dimmer sources offer fewer photons and the difference between regions of the spectrum that are partially absorbed and those that are fully unabsorbed becomes less apparent.

To quantify this trade-off between the decreasing trend in optical depth at lower redshift and the decreasing source flux density at high redshift, \citet{FOB} and \citet{Furlanetto:2006a} have related the minimum flux density $S_{min}$ of a source that could be used to detect with signal to noise $S/N$ a cloud of gas with optical depth $\tau$:
\begin{eqnarray} \label{eq:Smin}
S_{min} & = & 160 \textrm{ mJy} \left( \frac{S/N}{5} \frac{10^{-3}}{\tau} \frac{10^6 \textrm{ m}^2}{A_{eff}} \frac{T_{sys}}{400  \textrm{ K}} \right) \nonumber \\
             &     &  \times \left( \frac{1 \textrm{ kHz}}{\Delta \nu_{ch}} \frac{1 \textrm{ week}}{t_{int}} \right)^{1/2}.
\end{eqnarray}
The parameters $A_{eff}$ and $\Delta \nu_{ch}$ are the effective area of the telescope array and the channel bandwidth, here taken to be those plausible for the SKA, and $t_{int}$ is the integration time.  The system temperature, $T_{sys}$, is dominated by the Galactic synchrotron foreground at the frequencies of interest.  This can be approximated \citep{Carilli:2006} as 
\be \label{eq:systemp}
T_{sys} \sim 100 \left(\frac{\nu}{200 \textrm{ MHz}}\right)^{-2.8} \textrm{ K}.
\ee
This equation has been used as a proxy for the observability of the 21cm forest; in this work, we note that while a source above $S_{min}$ is required for a detailed study of the {\it structure} of the forest, it does not directly relate to the detectability of the presence of absorption or to the mean flux decrement that would be present in an IGM with mean optical depth $\tau$.  We elaborate on this point in Sections \ref{s:spectra} and \ref{s:stats} below.

To illustrate the high source fluxes required for detailed forest observations to be feasible, we include dashed lines corresponding to $S_{min}$ values in Figure \ref{sf:mean21od}, which plots the mean 21cm optical depth as a function of redshift for a range of values of $f_X$.  For each $f_X$ curve, if the mean optical depth line lies above the $S_{min}$ line for a given flux density, the signal-to-noise for that set of parameters should be at least 5 at that redshift.  In regions of the plot where the optical depth line lies below the $S_{min}$ line, a detection of the forest at signal-to-noise of 5 is not possible with the parameters of the observation that we have assumed.

Equation (\ref{eq:Smin}) illustrates the crucial trade-off in attempts to observe the 21cm forest.  Successful observations will require both a high optical depth to 21cm absorption and a population of very radio-loud sources.  The former pushes toward high redshift, while the latter is more feasible at low redshift.  The key issue, then, is whether there exist sufficiently radio-loud sources at sufficiently high redshift such that the IGM is still largely cold and neutral.  If radio-loud sources only occur within or in conjunction with massive halos that produce a large amount of ionizing radiation, or if early dense neutral regions are not well shielded, this trade-off becomes difficult to achieve.

Estimates of the abundance of radio-loud sources at high redshift are uncertain.  \citet{Haiman:2004} present an estimate of the number of radio-loud sources per square degree for a range of flux densities out to redshift $z=15$.  According to this estimate, in the redshift range $8 < z < 12$, there should be $\sim 2000$ sources above $ \sim 6$ mJy in the full sky.  However, these estimates depend strongly on assumptions about the radio-loud fraction and the spectral steepness of high-redshift sources.  \citet{Jiang:2007} have presented evidence that the radio-loud fraction may decrease with redshift, which would make discovery of appropriate 21cm forest background sources less likely.  In this work, we have shown that objects with 10-100 mJy at $z \gtrsim 8$ are most favorable as background sources.  If they exist, they may have already been discovered in surveys such as FIRST\footnote{http://sundog.stsci.edu/}.  \citet{Ivezic:2002} have found that $\sim$30\% of FIRST sources at 10-100 mJy have no detected optical counterpart within 3" (via SDSS).  If optical and infrared observations are carried out on these unidentified sources, some may be found to lie at high redshift, making them ideal candidates for observations of the 21cm forest.  We consider the issue of the abundance of background sources to be an open question, but we anticipate that it may be resolved soon with new observations and continued follow-up of known radio-loud sources.

In order to illustrate the qualitative properties of 21cm forest observations, we will assume a background source at high redshift with spectral properties similar to the powerful local radio source Cygnus A.  The intention in this assumption is not to assert that we would aim to carry out these observations against sources physically similar to Cygnus A at high redshift, or even that such sources necessarily exist, but rather to give a point of reference for the brightnesses and qualitative spectral features in our example calculations.

\begin{figure}
\centering
\subfloat[]
{\label{sf:Sweek}
\includegraphics[width=0.7\columnwidth,angle=-90]{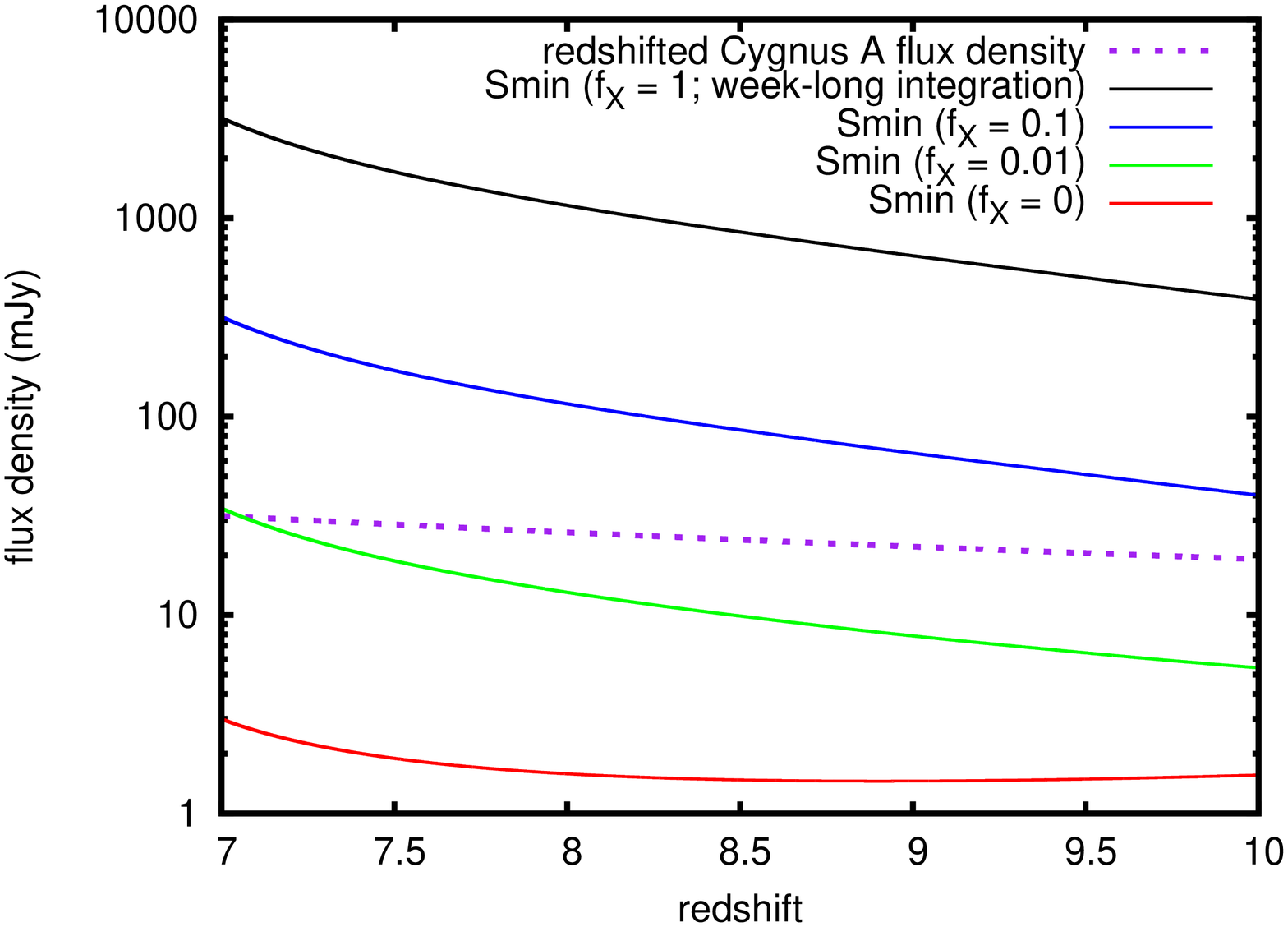}}

\subfloat[]{
\label{sf:Syear}
\includegraphics[width=0.7\columnwidth,angle=-90]{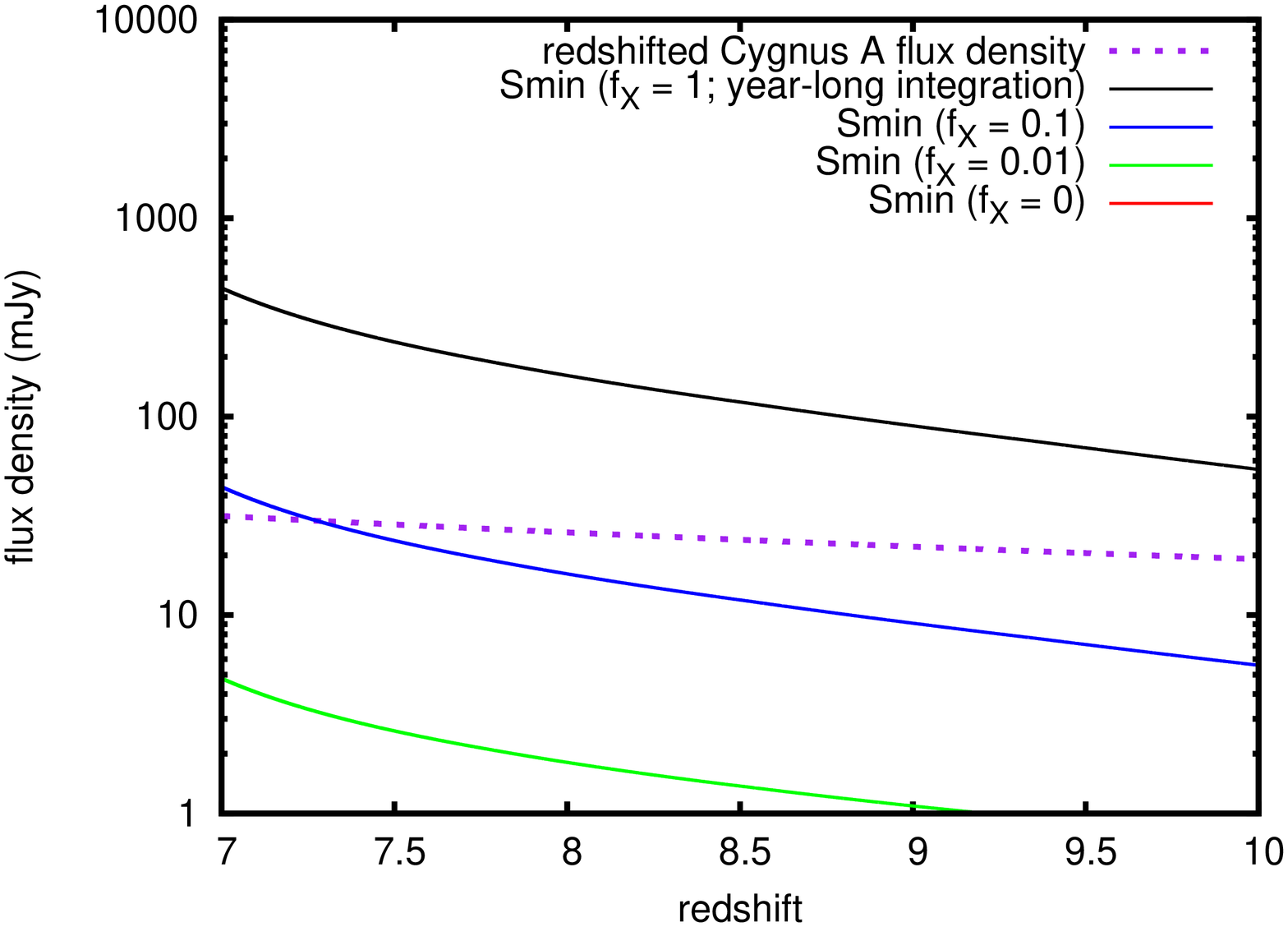}}

\centering
\caption{\label{f:Smincurves}  Minimum flux density curves for $S/N \gtrsim 5$ detection of the 21cm forest for $f_X=(0, 0.01, 0.1, 1)$ (solid lines), with the redshifted 21cm flux density of a Cygnus A-like source also plotted as a function of redshift (dashed line).  The instrument is assumed to be SKA with channel width of 1 kHz and integration time of \subref{sf:Sweek} 1 week and \subref{sf:Syear} 1 year.  In \subref{sf:Syear}, the $f_X=0$ line is below the limits of the plot.}
\end{figure}

Figures \ref{sf:Sweek} and \ref{sf:Syear} use the spectrum of our hypothetical Cygnus A-like source to illustrate in a different way the trade-off between optical depth and source brightness.  For $f_X$=(0, 0.01, 0.1, 1), we plot as a function of redshift the minimum flux density $S_{min}$ required for an SKA-like instrument to detect the 21cm forest at signal-to-noise of 5.  We assume frequency resolution $\Delta \nu_{ch} = 1$ kHz and integration times $t_{int}$ of one week and one year, respectively.  On the same axes, we also plot the flux density of a Cygnus A-like source as it would be observed at that redshift, at an observing frequency corresponding to the rest-frame 21cm line.  In this plot, if the $S_{min}$ curve for a particular value of $f_X$ is below the Cygnus A flux density curve, then a source similar to Cygnus A would be sufficiently bright to allow the detection.  We see that for a week-long integration, detection is possible only if $f_X \lesssim 0.01$, even for a source existing at redshift as high as $z=10$.  For the same observation with a year-long integration, the example radio source would be sufficiently bright to allow detection for $f_X=0.1$, but the $f_X=1$ scenario still does not provide sufficiently high optical depth.  A similar exercise for $\Delta \nu_{ch}=100$ kHz (which reduces $S_{min}$ by an order of magnitude in each case) shows that for $f_X=1$, a week-long integration would still be insufficient, but a year-long integration could detect the forest at high significance against a Cygnus A-like source at $z \gtrsim 7$.

\section{Simulated spectra}
\label{s:spectra}

\begin{figure}
\centering
\subfloat[]
{\label{sf:fx1spectrum}
\includegraphics[width=0.7\columnwidth,angle=-90]{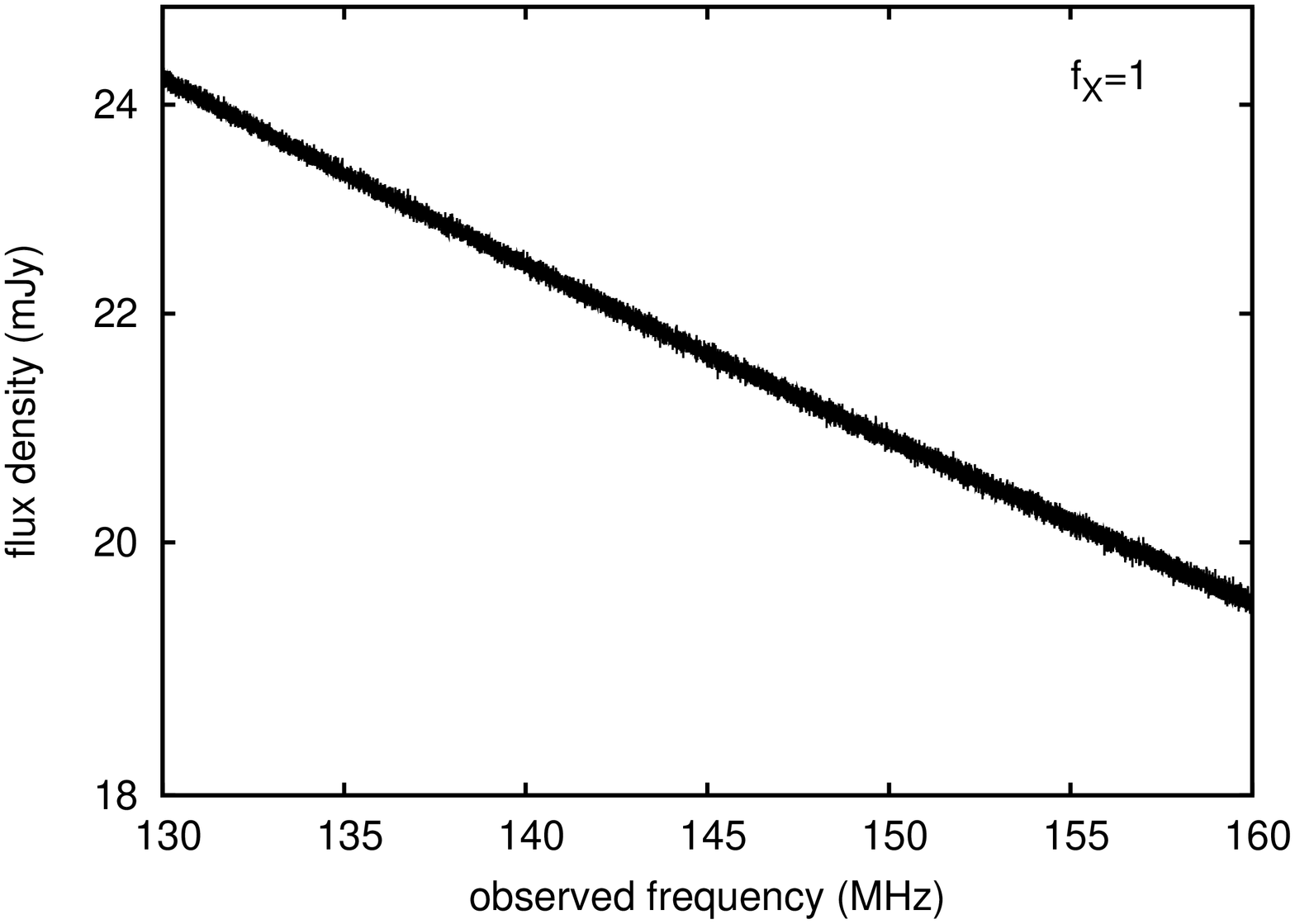}}

\subfloat[]{
\label{sf:fx01spectrum}
\includegraphics[width=0.7\columnwidth,angle=-90]{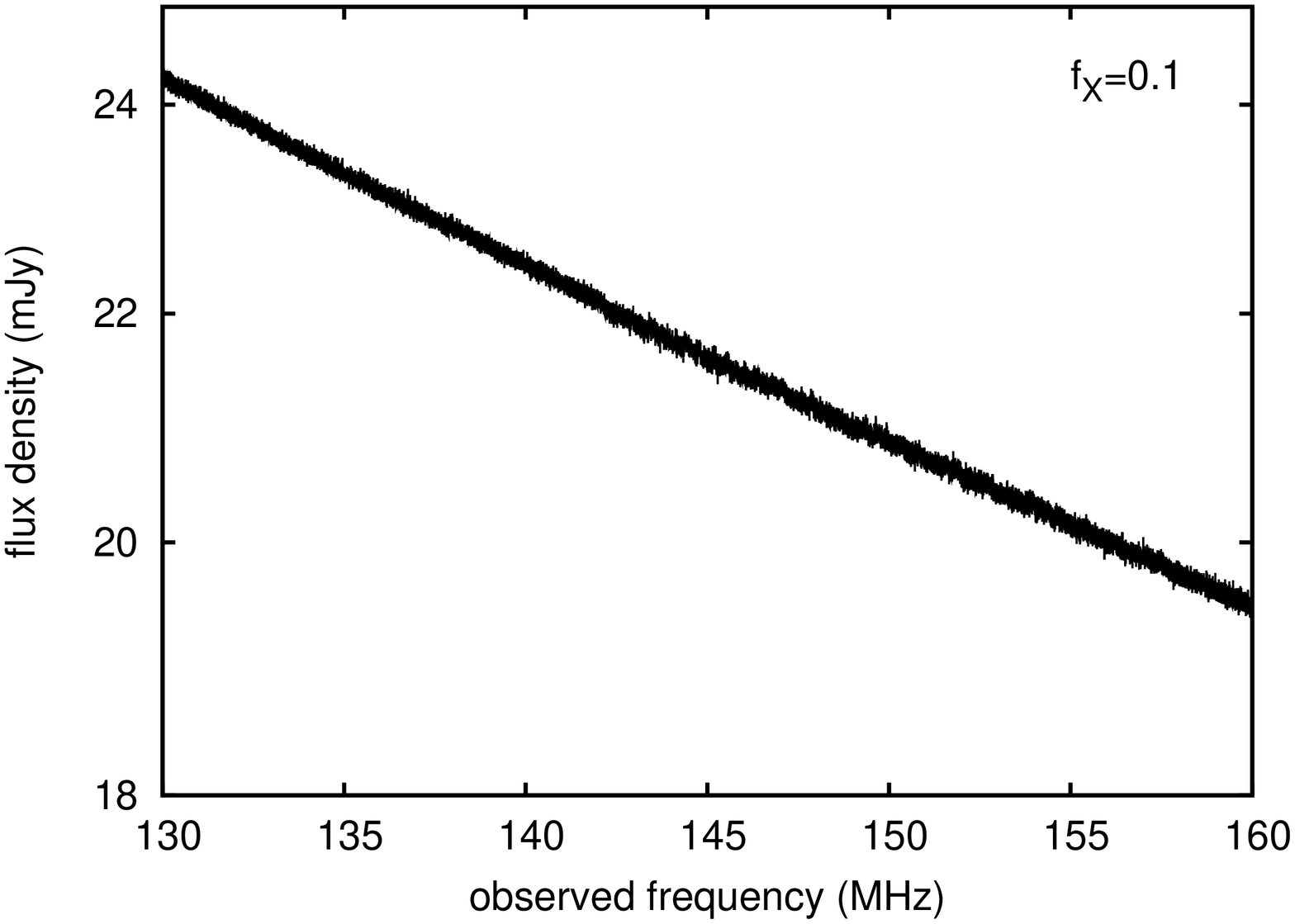}}

\subfloat[]{
\label{sf:fx001spectrum}
\includegraphics[width=0.7\columnwidth,angle=-90]{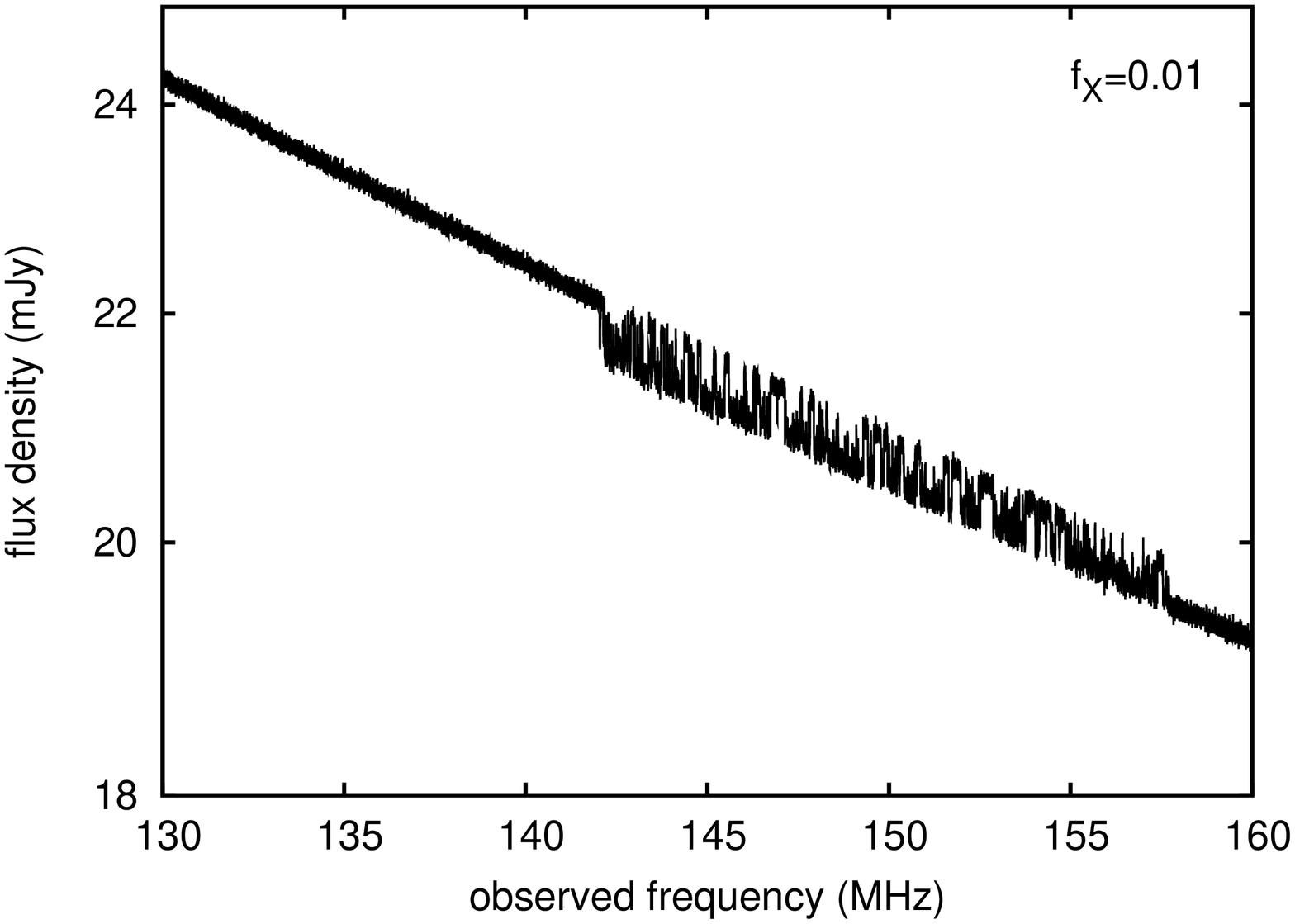}}

\centering
\caption{\label{f:fxspectra}  Simulated spectrum for 21cm absorption against the spectrum of a radio source at $z=9$ with spectral properties similar to that of Cygnus A.  The x-ray efficiency factors in these simulations are \subref{sf:fx1spectrum} $f_X=1$, \subref{sf:fx01spectrum} $f_X=0.1$ and \subref{sf:fx001spectrum} $f_X=0.01$.  Measurement noise appropriate to observation with SKA, with $\Delta \nu_{ch}=1$ kHz and $t_{int}=1$ week is included. In this simulation, overdense neutral regions (e.g., minihalos or protogalaxies) along the line of sight are not included. }
\end{figure}

In order to simulate a realistic 21cm forest measurement, we have produced synthetic spectra for absorption against a hypothetical Cygnus-A-like source placed at $z=9$, with added noise appropriate to an observation with the SKA, based on noise estimates from \citet{Carilli:2004}.  For a spectrum with 1 kHz frequency resolution and a week-long integration, we assume Gaussian noise at a level of $\sigma \approx 41 \mu$Jy.  In Figures \ref{sf:fx1spectrum}, \ref{sf:fx01spectrum} and \ref{sf:fx001spectrum}, we plot the 21cm absorption spectrum with noise included for x-ray heating with $f_X=$ 1, 0.1, and 0.01, respectively.

Although the absorption due to the 21cm forest is marked in the $f_X=0.01$ scenario, it is difficult to see by eye in the $f_X=0.1$ spectrum and is visually indistinguishable from noise for $f_X=1$.  In this simulation, we have included only mild overdensities with $\delta \sim 1$, and the prescription used to simulate the ionization state of the gas results in an ionization fraction correlated with density.  Therefore there are no highly overdense neutral regions included here.  

In works by other authors [e.g., \citet{Carilli:2004,Xu:2009}], dense neutral regions produce sharp absorption spikes in the 21cm forest spectrum. 
Our simulations do not include these regions (protogalaxies or minihalos -- the progenitors of damped Lyman-$\alpha$ systems).  As discussed in Section \ref{s:numsims} above, the estimates of the number and character of such absorbers are highly uncertain, and their dependence on the state of the mean IGM (which is our main interest here) is not straightforward.  However, we also recognize that the inclusion of such systems might be important in the analysis of the spectra. Therefore, we have developed a stochastic method to produce absorption features ``by hand'' which we then insert into the spectra to simulate the effect of these narrow absorption lines.  We produce $N$ spectral lines [where $N=(5,10,50,100,200,300,400,500)$] by drawing from a distribution of optical depths based on the calculation in \citet{Furlanetto:2006a} and a linewidth of 2-3 kHz.  After including instrumental noise, we randomly place these lines in the spectra in the non-uniform simulation region between 142 and 148 MHz (corresponding to $z=8-9$).  In order to take into account the fact that small halos would likely be photoevaporated in highly ionized bubbles, we throw out any lines that are assigned to high-ionization regions.  The resulting spectra are not meant to be quantitative predictions of future observational results, but rather to provide a plausible basis for the determination of the study's feasibility and specifically to show how the presence of absorption lines would affect estimates of $f_X$ based on the study of the spectra.

In Figure \ref{f:lines_spectra} below, we show simulated spectra for $f_X=0.1$ with 100 and 500 absorption lines added, in \ref{sf:fx01_100lines_spectrum} and \ref{sf:fx01_500lines_spectrum}, respectively.  Note that since lines randomly assigned to ionized regions are not included, the number of lines appearing in the spectrum is somewhat smaller than the number of added lines (around \nicefrac[]{3}{4} of the lines remained in these cases).  Comparing Figure \ref{f:lines_spectra} with Figure \ref{sf:fx01spectrum}, we see that when absorption lines are included, while the flux decrement and bubble structure are still present, the most prominent feature of the spectrum becomes the forest of absorption lines.

\begin{figure}
\centering
\subfloat[]
{\label{sf:fx01_100lines_spectrum}
\includegraphics[width=0.7\columnwidth,angle=-90]{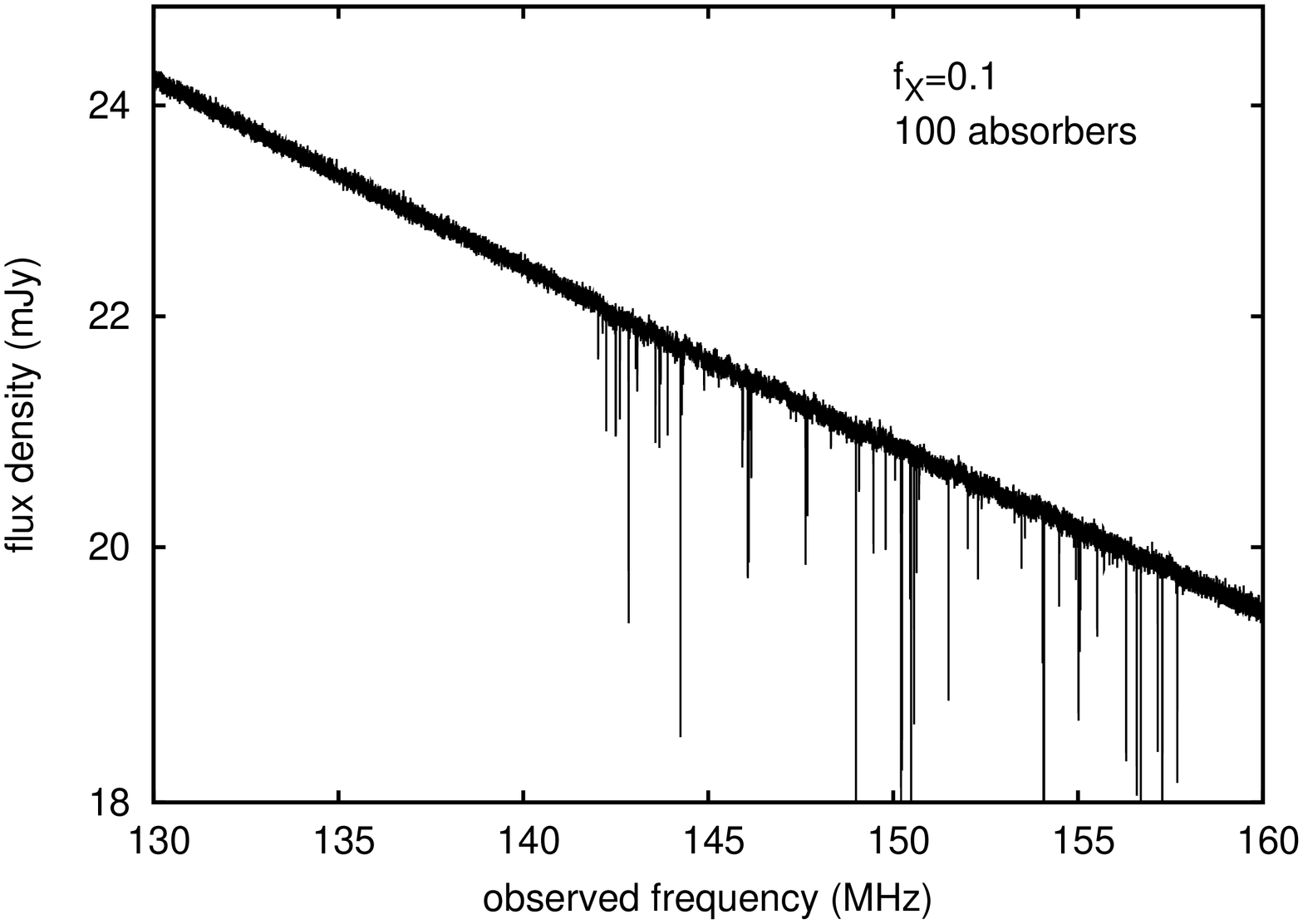}}

\subfloat[]{
\label{sf:fx01_500lines_spectrum}
\includegraphics[width=0.7\columnwidth,angle=-90]{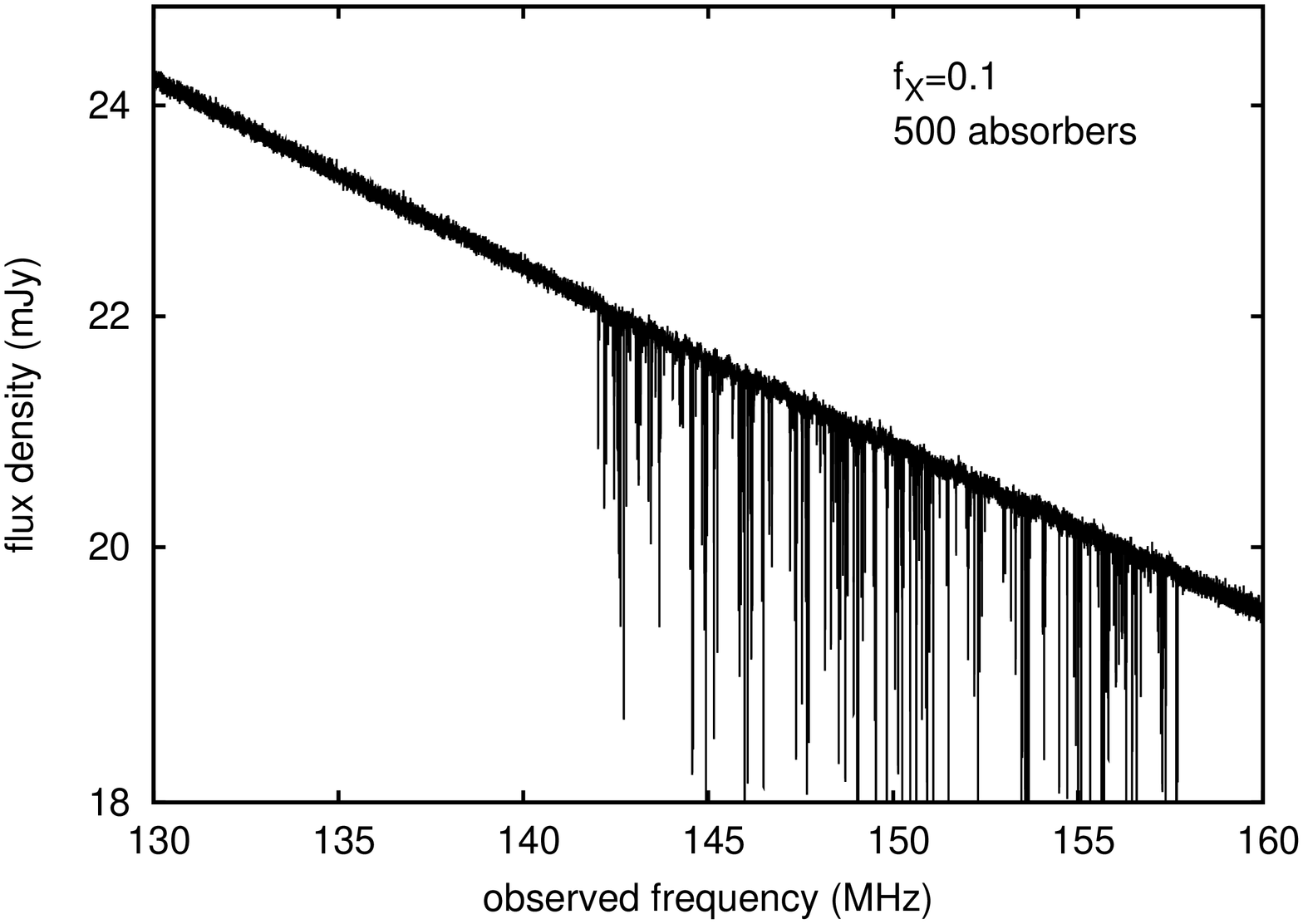}}

\centering
\caption{\label{f:lines_spectra}  Simulated spectrum for 21cm absorption against the spectrum of a radio source at $z=9$ with spectral properties similar to that of Cygnus A, with \subref{sf:fx01_100lines_spectrum} 100 and \subref{sf:fx01_500lines_spectrum} 500 absorption lines added to simulate the presence of overdense neutral regions along the line of sight.  The x-ray efficiency factor in these simulations is $f_X=0.1$.  Observational parameters are as in Figure \ref{f:fxspectra} above.} 
\end{figure}

\section{Statistical detection methods}
\label{s:stats}

Here we present two methods to derive an estimate of $f_X$ from the 21cm forest spectrum.  In the first method, we show that the structure of transmission regions in the spectrum (due to photoionized regions in the line of sight) leads to an increase in the spectrum's variance in the absorbed region.  The relative flux between the absorption and transmission regions is determined by the mean optical depth, which depends on the gas temperature, which itself depends on $f_X$.  Therefore $f_X$ can be determined by the degree to which the spectrum's variance increases in the region blueward of the redshifted 21cm transition.  This method is most appropriate when there are no strong absorption lines in the spectrum.

The second method can be used even when many strong absorption lines are present.  In this method, the spectrum is ``normalized'' (source-subtracted) and divided into the absorbed and unabsorbed regions (blueward and redward of the 21cm transition, respectively).  The flux distribution in each region is fit to a Gaussian, from whose shape we can derive properties of the IGM.  The anomalously low fluxes in pixels corresponding to narrow absorption lines cause them to be outliers in the flux distribution, which effectively allows us to subtract them from the signal of the ``typical'' IGM that would be seen if no lines were present.  Thus they do not hinder our ability to determine the properties of the IGM, which is our main interest in this work.

In an actual observation, the intrinsic spectrum of the source would not be known -- it might instead be approximated by a powerlaw of unknown slope. Bright radio foregrounds would also complicate the subtraction. Since the frequency range of the measurement is much smaller than the central frequency ($\delta \nu / \nu \ll 1$), in practice it may be possible to accurately extrapolate the foreground and continuum component from the unabsorbed side to the absorbed side of the spectrum.  We therefore consider our results utilizing the source-subtracted (normalized) spectrum to be a best-case scenario.  In particular, a flux decrement measurement (see Section \ref{s:gauss}) over a wide band would be highly dependent on accurate foreground and source subtraction.  However, estimates of the variance discussed in Sections \ref{s:variance} and \ref{s:gauss} below are not subject to this continuum uncertainty, as the width of the spectrum's flux distribution should be robust even if its normalization is not known precisely.

In Sections \ref{s:variance} and \ref{s:gauss} below, we describe these statistical detection methods in more detail and show how they can be used to study the thermal history of the IGM in cases where the features of the 21cm forest cannot be studied at high signal-to-noise.

\subsection{ Variance increase in absorbed region}
\label{s:variance}

As can be seen in the $f_X=0.01$ case without absorption lines included (Figure \ref{sf:fx001spectrum}), the most striking features of the spectrum  are the step-like transitions between transmission through photoionized bubbles and absorption in largely neutral regions.   The detection limits discussed in Section \ref{s:sourcepop} can be stated in terms of the minimum flux density $S_{min}$ of a background source that would allow a clear detection of an absorption feature.  Individual absorption features can be seen clearly in Figure \ref{sf:fx001spectrum}, as can the ``DC offset'' or flux decrement in the absorption region, but in Figure \ref{sf:fx01spectrum} the flux decrement is less apparent, and in Figure \ref{sf:fx1spectrum} the offset cannot be seen by eye.  However, one can see in Figure \ref{sf:fx01spectrum} (and to a much greater extent in Figure \ref{sf:fx001spectrum}) that the variance (or apparent noisiness) of the spectrum increases at the onset of 21cm absorption, around 142 MHz.  This increase in variance, discussed by \citet{Carilli:2004}, may be used to detect the 21cm forest statistically in spectra for which the signal-to-noise is low. 

Using our synthetic 21cm forest spectra, we investigate the effectiveness of the statistical detection technique.  After smoothing the spectrum to a given frequency resolution $\Delta \nu_{ch}$, we add Gaussian noise to each pixel in frequency space and then subtract the input source spectrum.
The source-subtracted spectrum is then broken into regions of width $w$, and in each of these windows, the variance of the spectrum is calculated.  For a spectrum with no absorption, as is seen in frequencies redward of the source redshift, the variance should be equal to the variance of the input Gaussian noise, $\sigma^2$.  Blueward of the frequency of the onset of absorption, the variance increases as the absorption level varies due to density and ionization-state fluctuations in the intervening IGM.  The largest variances occur between regions of maximal absorption (from substantially neutral clumps of gas) and regions of no absorption (photoionized bubbles).  Therefore a very small $w$ (of order the frequency resolution $\Delta \nu_{ch}$) will pick up little variance above that of the Gaussian noise, since the variance within the window comes primarily from the signal variance due to noise.  Windows of order the size of photoionized bubbles in frequency space will have the maximum variance, as they will generally cut across regions of maximal and minimal absorption.  As the window size increases further, the variance in the forest region will eventually level off as a large number of bubbles are averaged over and the mean optical depth drops at high frequencies.

To illustrate this, we calculate the mean variance in windows of different widths, $w$, on either side of the onset of 21cm absorption.  For absorption against a hypothetical radio-loud source at $z=9$, we can split the spectrum roughly into the pre-absorption and post-absorption regions by comparing the signal variance in the region from 130 to 140 MHz to the signal variance in the region from 140 to 150 MHz. The mean signal variance $\langle \sigma_i^2 \rangle$ in a window of width $w$ is the mean of
\be
\sigma_i^2 = \displaystyle \sum \limits_{\nu = \nu_0}^{\nu_0+w} (F_{N,\nu} - \langle F_{N,i} \rangle)^2
\ee
where $\nu_0$ is the starting frequency in window $i$, $F_{N,\nu}$ indicates the source-subtracted flux at frequency $\nu$ and $\langle F_{N,i} \rangle$ is the mean source-subtracted flux across window $i$.  The window sizes considered range from 50 kHz to 5 MHz, and we look specifically at models with $f_X=0.1$ and $f_X=1$ in order to show how the variance increase can improve the detectability of the 21cm forest in cases without a strong DC offset.

\begin{figure}
\centering
\subfloat[]
{\label{sf:01noisebins}
\includegraphics[width=0.7\columnwidth,angle=-90]{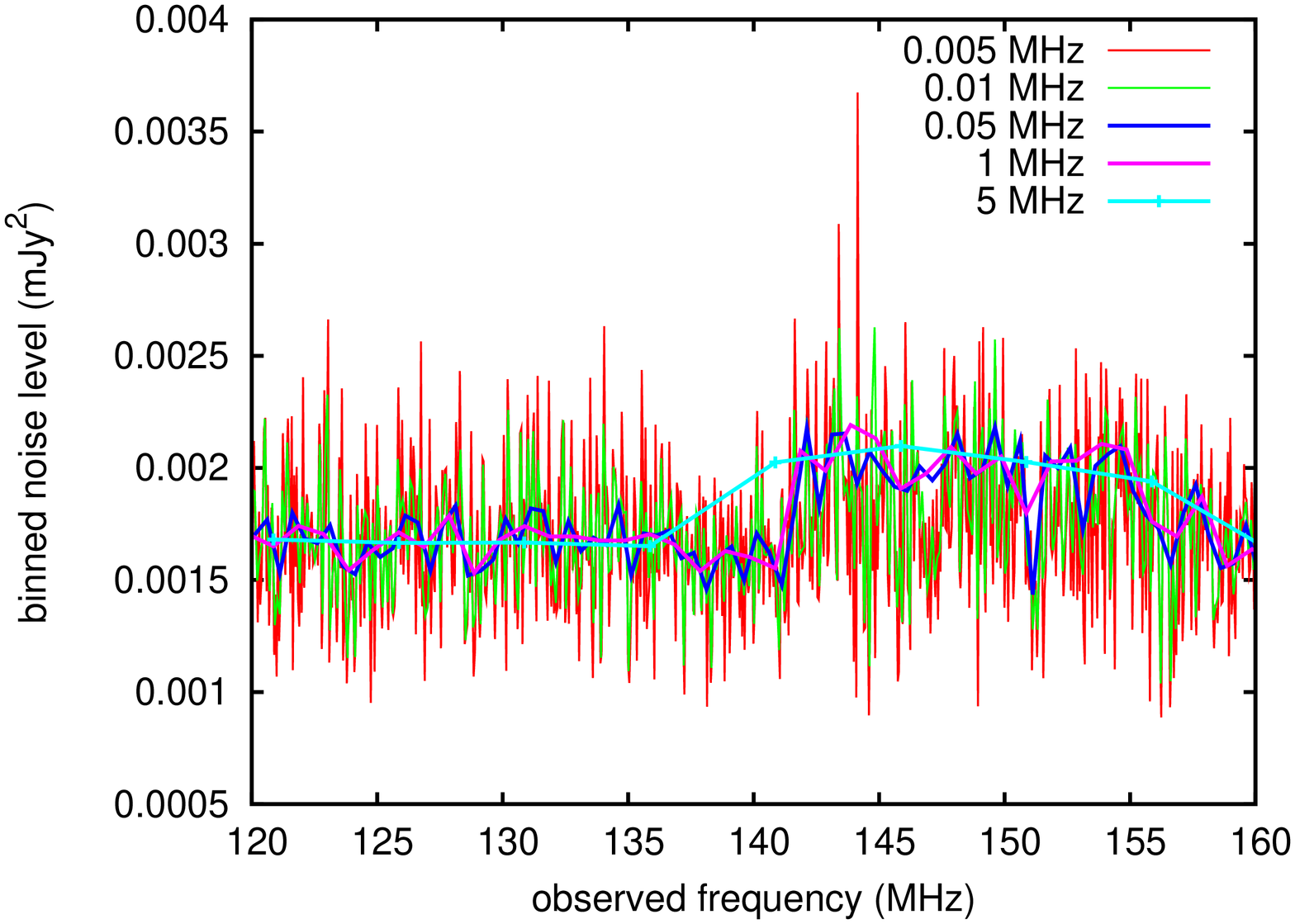}}

\subfloat[]{
\label{sf:1noisebins}
\includegraphics[width=0.7\columnwidth,angle=-90]{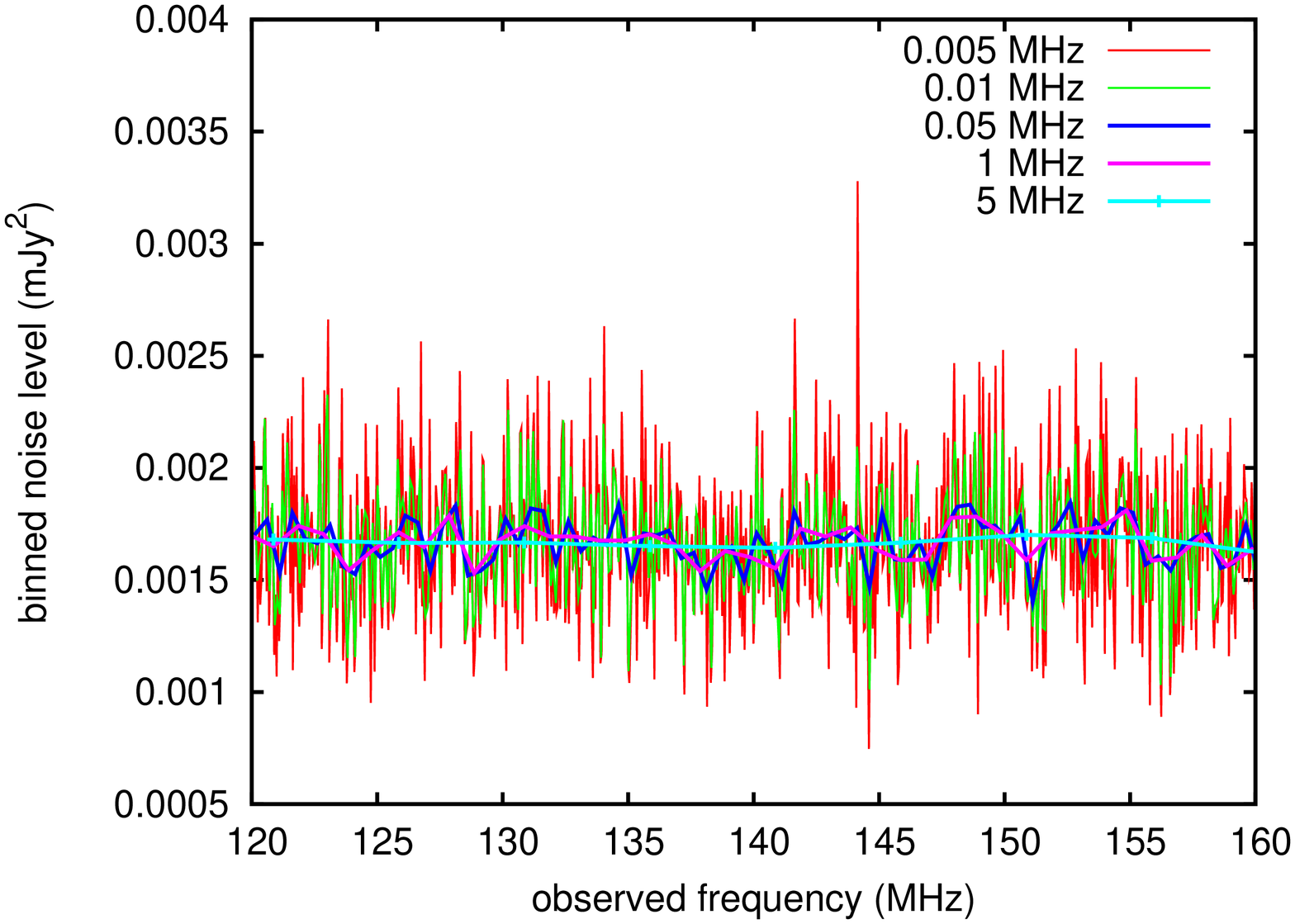}}

\centering
\caption{\label{f:noisebins}  Mean binned signal variance as a function of frequency for windows with $w=$ (50 kHz, 100 kHz, 500 kHz, 1 MHz, 5 MHz), for \subref{sf:01noisebins} $f_X = 0.1$ and \subref{sf:1noisebins} $f_X = 1$.  No absorbers are included in this simulation.}
\end{figure}

In Figures \ref{sf:01noisebins} and \ref{sf:1noisebins}, we plot the mean binned signal variance as a function of frequency for windows of different sizes, for $f_X=0.1$ and $f_X=1$, respectively.  In the $f_X=1$ case, the increase in variance at the onset of absorption is not apparent by eye for any of the window sizes.  In the $f_X=0.1$ case, on the other hand, the effect of the photoionization bubbles is clear in the increase in variance.  For small window sizes ($w = 50-100$ kHz), the increase in variance at the onset of absorption is difficult to discern, as the scales probed are too small to pick up typical photoionization bubbles, and the variance is similar to that of the Gaussian noise ($\sigma^2 \approx 1.7 $ mJy$^2$).  For larger window sizes ($w = 0.5-5$ MHz), however, there is a clear increase in the mean variance after the onset of absorption at $\nu \approx 142$ MHz.  It is important to note that the curves for the larger window sizes are not simply a smoothed version of the curves for smaller windows: except in the pre-absorption region where there is no 21cm signal, the 5 MHz curve would fail as a fit to the 50 or 100 kHz curves.  This is because the larger windows are seeing more of the photoionization bubble structure, whereas the smaller windows are largely missing it and picking up primarily the Gaussian signal variance on top of the (relatively weak) optical depth variations due to mild overdensities and regions of fractional ionization.
 
\begin{figure}
\begin{center}
\resizebox{\columnwidth}{!}{\includegraphics[angle=-90]{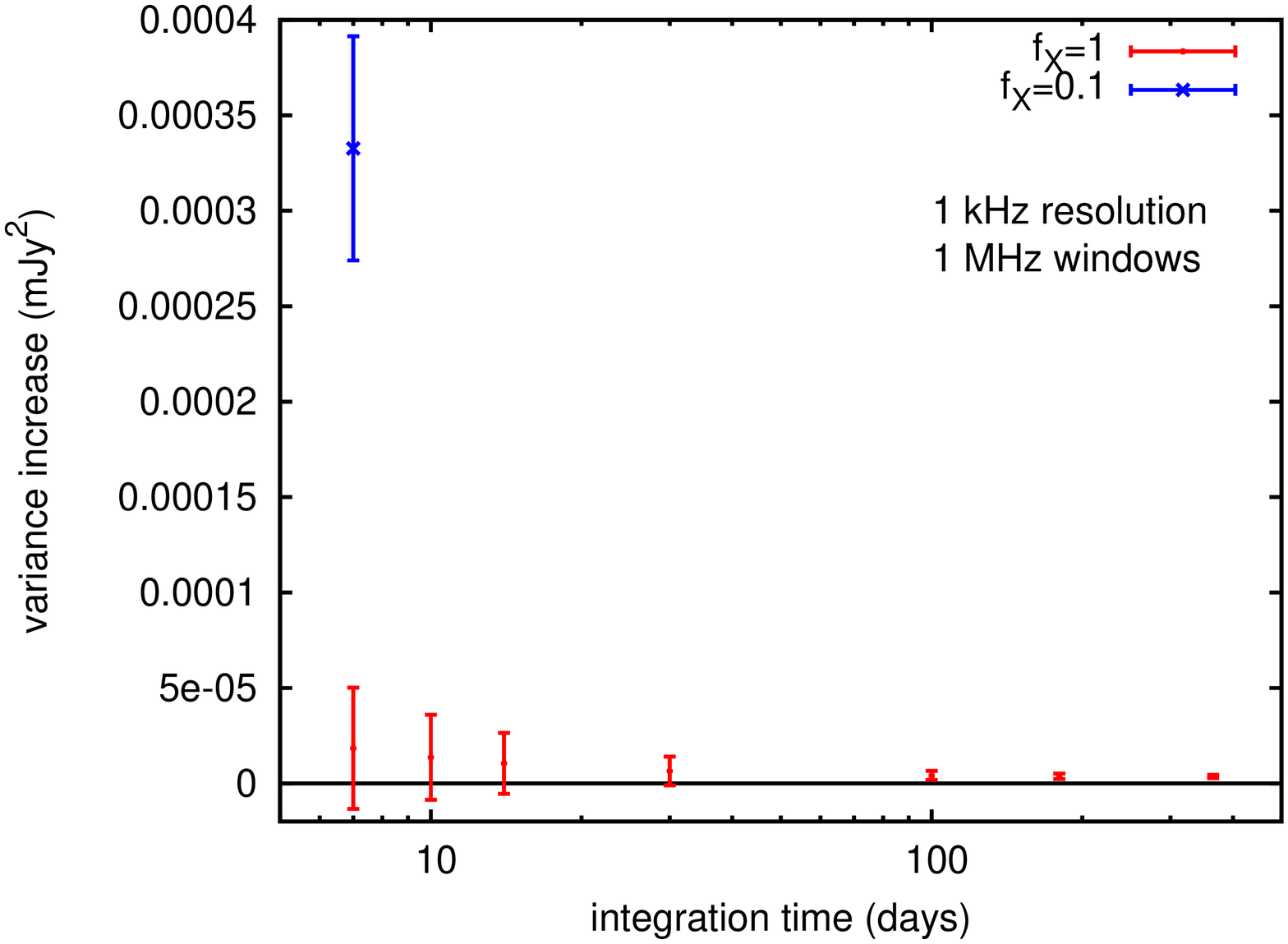}}
\end{center}
\caption{Difference in mean variance between unabsorbed and absorbed regions as a function of integration time for a window size of 1 MHz, with $f_X=1$ and no lines added (red symbols with errorbars).  Also included, for reference, is the difference in mean variance for $f_X=0.1$ with a window size of 1 MHz, for an integration time of 1 week (blue cross with errorbars).}
\label{f:new_int_time}
\end{figure}

For the purpose of this calculation, we have defined the unabsorbed region of the spectrum to be the frequency range 130-140 MHz and the absorbed region 140-150 MHz [though in an actual observation, if we know the redshift with high accuracy we can obtain greater precision by using 1420 MHz $\times (1+z)$ as the dividing line].  In each region, we take the mean of the variance and calculate the difference between the means in the two regions. In Figure \ref{f:new_int_time}, we plot this difference as a function of integration time for a window of size 1 MHz and $f_X=1$.  For comparison, we also include the difference for $f_X=0.1$ at an integration time of 1 week.  This plot shows that for small integration times, the variance method still will not allow us to detect the forest against a Cygnus-A-like source for $f_X=1$.  However, increasing the integration time can allow a significant detection even for this case. (Compare with Figure \ref{sf:Syear}.)  This suggests that even if $f_X$ is high and very radio-loud sources are not available for high signal-to-noise in spectral measurements of the 21cm forest, a detection of absorption (and therefore a constraint on $f_X$) may be possible with extended observations.

\subsection{Gaussian fitting}
\label{s:gauss}
 
In spectra with numerous strong absorption lines, the signal variance is dominated by the lines, making the statistical method described in Section \ref{s:variance} above more difficult.  In order to take the variance due to lines into account, one would have to have a detailed model of the way in which $f_X$ affects the number and character of absorption lines.  Since such models are still highly uncertain, we developed a method to effectively remove the absorption lines from the spectrum in order to better characterize the thermal properties of the mean IGM.

Our method uses Gaussian fitting essentially as a filter to remove highly non-Gaussian elements (i.e., narrow absorption lines).  First, we source-subtracted the spectra as discussed in Section \ref{s:variance}.  For a spectrum with no absorption lines added, this results in a ``normalized'' spectrum in which the unabsorbed region consists of Gaussian noise with a mean of zero and the absorbed region contains both Gaussian noise and a mildly non-Gaussian distribution of absorption and transmission regions with a mean below zero -- this mean is the flux decrement. We reiterate that this depends strongly on how perfectly the source spectrum can be subtracted in practice.

Let $F_0(\nu)$ be the flux density of the source as a function of frequency.  In a pixel at some frequency $\nu_0$, the observed flux is
\be
F(\nu_0) = F_0(\nu_0) T(\nu_0) \approx F_0(\nu_0)[1-\tau(\nu_0)]
\ee
where $T(\nu)$ is the transmission fraction at frequency $\nu$ and $\tau(\nu) \ll 1$ is the corresponding optical depth.  If we source-subtract the spectrum, we have
\be
F_N(\nu_0) = -F_0(\nu_0) \tau(\nu_0)
\ee
where $F_N$ indicates the source-subtracted (normalized) flux.  For the unabsorbed region in which $\tau=0$ (i.e., redward of the redshifted 21cm frequency), $F_N=0$ on average, the only variation being due to instrumental noise, which we assume here to be Gaussian.  We denote the average flux of the unabsorbed region on the left-hand side of the spectrum $F_L$ and the average flux of the absorbed region on the right-hand side $F_R$.  For a small frequency range around the 21cm transition, $\langle F_L \rangle = 0$ and 
\be \label{eq:taufromfluxdec}
\langle F_R \rangle = - \langle F_{0,R} \rangle \langle \tau_R \rangle .
\ee
This defines the average {\it flux decrement} $\Delta F_N$ of the absorption region:
\be
\Delta F_N = \langle F_R \rangle
\ee
Note that we have chosen a convention in which the flux decrement is a negative quantity -- strictly the (negative) distance from $F_N=0$ rather than the (positive) distance below $F_N=0$.

\begin{figure}
\centering
\subfloat[]
{\label{sf:fx01_nolines_left}
\includegraphics[width=\columnwidth,angle=0]{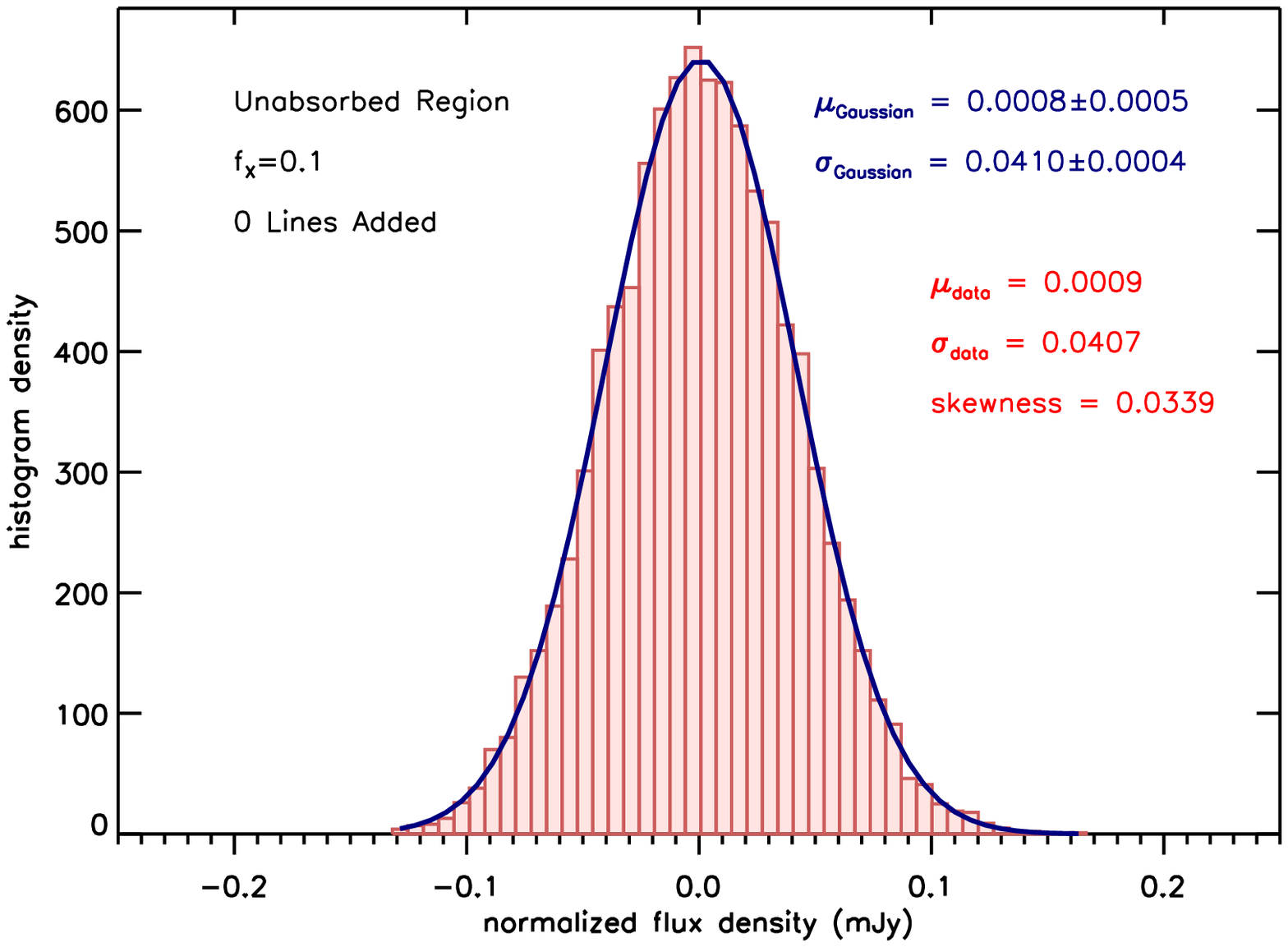}}

\subfloat[]{
\label{sf:fx01_nolines_right}
\includegraphics[width=\columnwidth,angle=0]{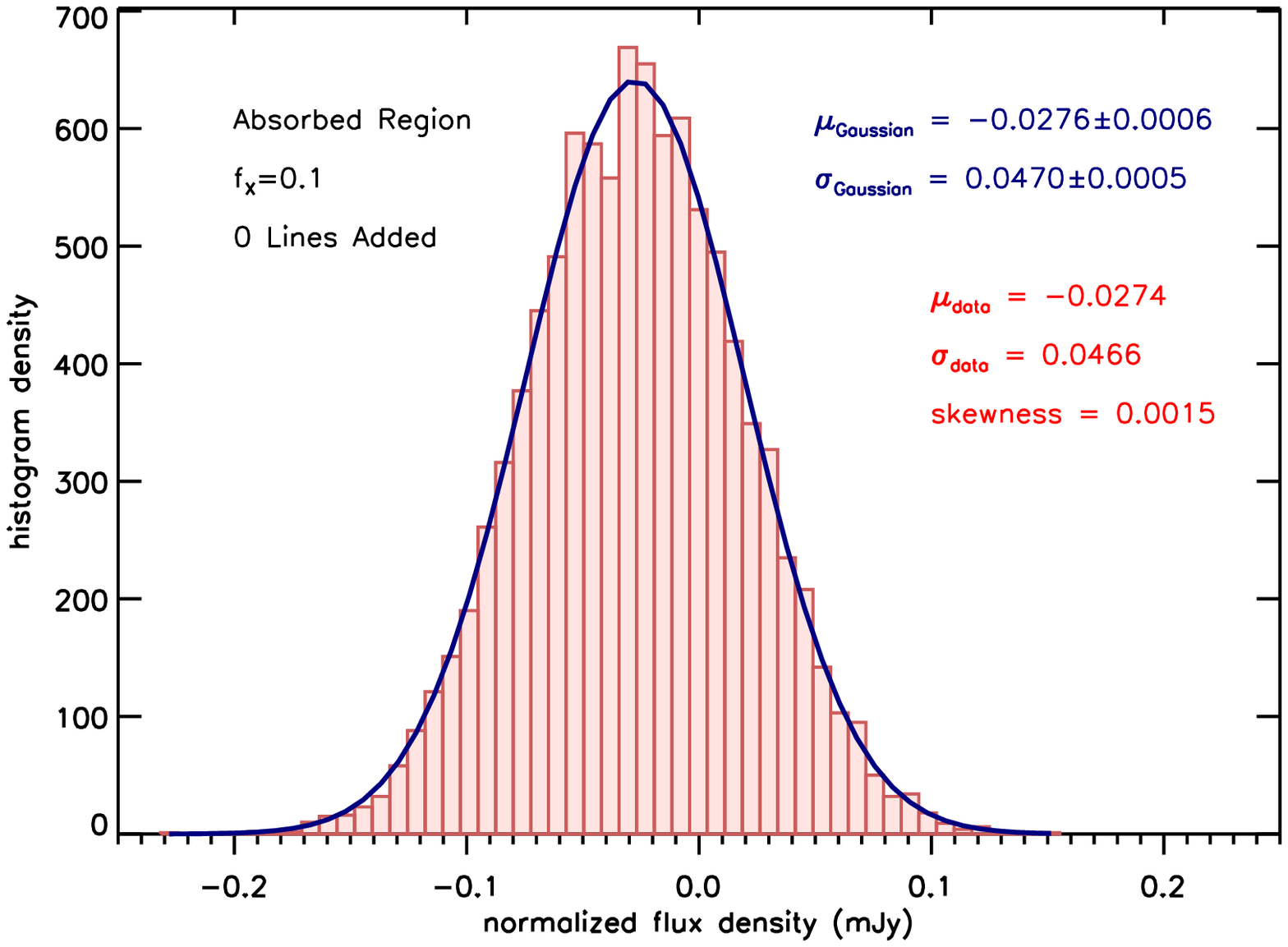}}

\subfloat[]{
\label{sf:fx01_100lines_right}
\includegraphics[width=\columnwidth,angle=0]{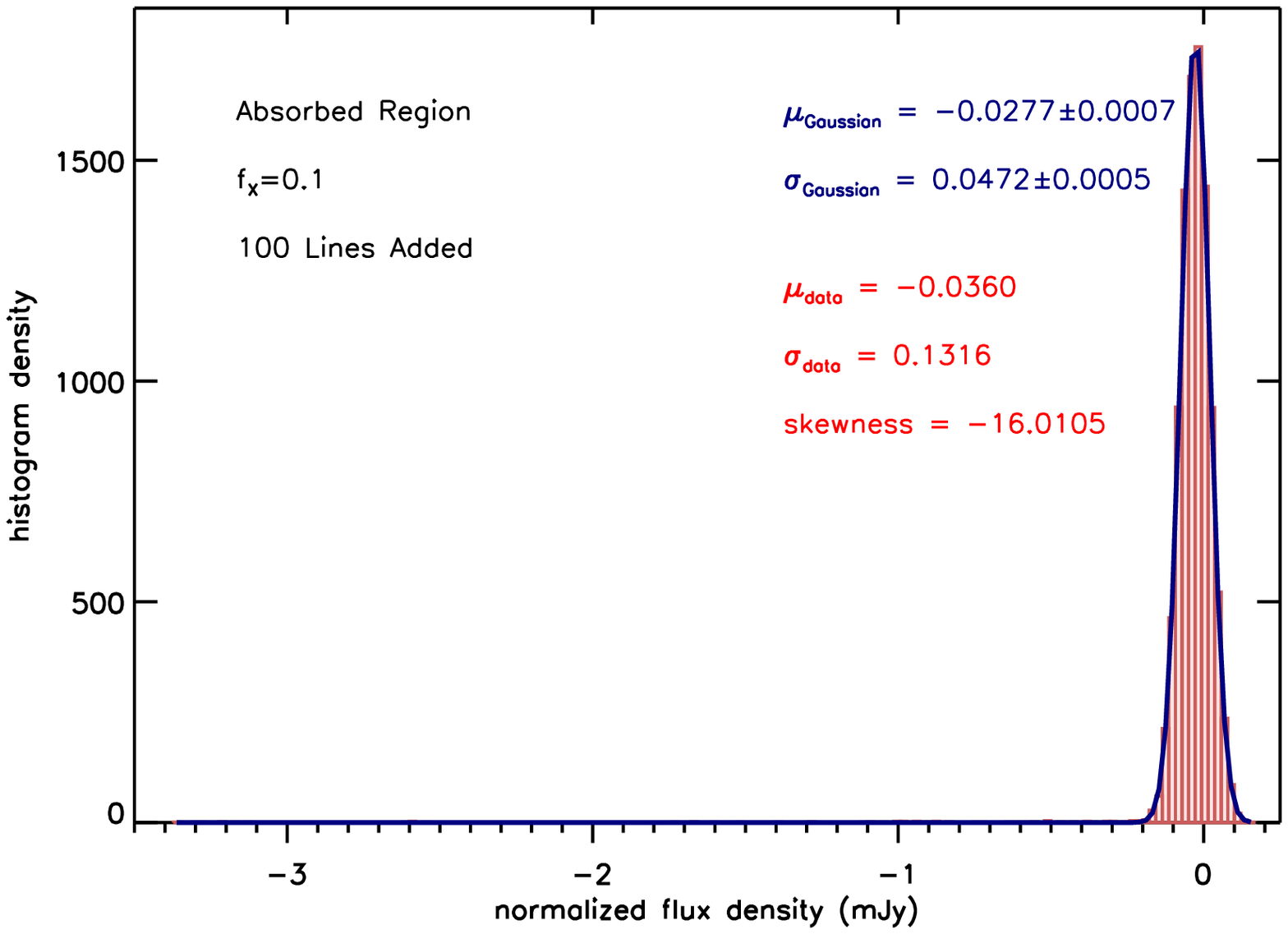}}

\centering
\caption{\label{f:gaussianhists} Gaussian fit (blue curve) to histogram of normalized flux densities from $f_X=0.1$ spectrum for the \subref{sf:fx01_nolines_left} unabsorbed and \subref{sf:fx01_nolines_right} absorbed region of the spectrum with no absorption lines added, as well as the \subref{sf:fx01_100lines_right} absorbed region with 100 lines added.  The mean and standard deviation of both the spectrum points (``data'') and Gaussian fit are included on the plots, as is the skewness of the spectrum points in each region.  Note that in \subref{sf:fx01_100lines_right}, the x-axis range has been extended to reflect the fact that the histogram extends to large negative values of the normalized flux, though the number of points in each low-flux bin is too small to be seen in the plot.}
\end{figure}

If no lines are present in the spectrum, the mean normalized flux in the absorbed region gives the flux decrement, which can be related in a straightforward way to the optical depth and, consequently, $f_X$.  However, if absorption lines are present in the spectrum, they can distort the measured average flux decrement, making it difficult to straightforwardly relate it to the mean $\tau$ in the IGM.  Similarly, absorption lines dominate the signal variance.  Gaussian fitting solves this problem.  Instead of calculating the mean or standard deviation of the flux distribution itself, we make a histogram of the flux distribution of each region (absorbed and unabsorbed) and fit this to a Gaussian.  The mean and standard deviation of the Gaussian fit are close to those that would be found if the lines were not included in the spectrum, since the lines constitute non-Gaussian outliers that, due to their narrowness and the large region of the spectrum considered, have little weight in the fitting calculation.  In Figure \ref{f:gaussianhists}, we show example histograms and Gaussian fits for the unabsorbed region of the spectrum as well as the absorbed region with 0 and with 100 lines added.  In each plot, we also include: $\mu_{\textrm{Gaussian}}$ and $\sigma_{\textrm{Gaussian}}$, the mean and standard deviation of the Gaussian fits, and $\mu_{\textrm{data}}$ and $\sigma_{\textrm{data}}$, the mean and standard deviation of the simulated spectrum ``data.'' We also include the skewness of the simulated spectrum points.  The non-Gaussian outliers have a strong influence on the skewness of the distribution; in this case, a large (negative) skewness indicates the presence of many absorption lines.  In Figure \ref{sf:fx01_100lines_right}, the range of the plot extends far to negative flux values to include the bins in the histogram corresponding to the absorption lines.  As is evident in this plot, while the absorption lines extend to low fluxes, their distribution is such that they cannot be seen by eye in the histogram and they do not significantly affect the location of the Gaussian fit.

\begin{figure}
\begin{center}
\resizebox{\columnwidth}{!}{\includegraphics[angle=0]{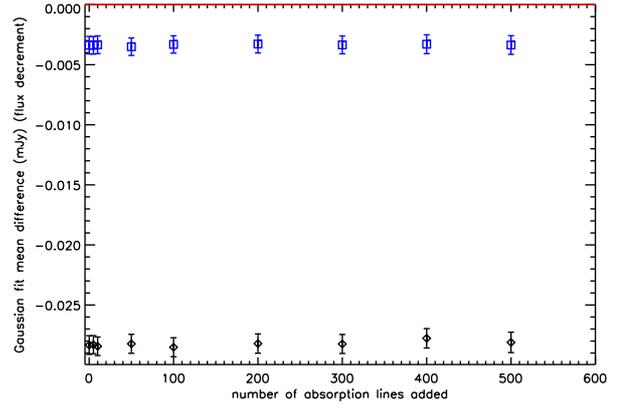}}
\end{center}
\caption{Difference between the Gaussian mean in the absorbed region and unabsorbed region of the spectrum (flux decrement) for $f_X=1$ (blue squares) and $f_X=0.1$ (black diamonds), plotted against the number of absorption lines added to the spectrum.  Error bars are derived from a bootstrap resampling of each spectrum.}
\label{f:meansplot}
\end{figure}

\begin{figure}
\begin{center}
\resizebox{\columnwidth}{!}{\includegraphics[angle=0]{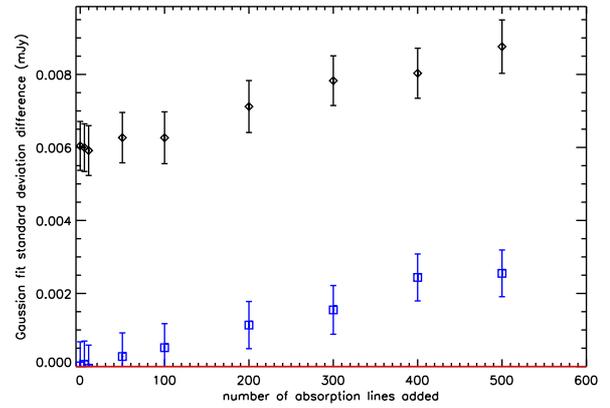}}
\end{center}
\caption{Difference between the Gaussian standard deviation in the absorbed region and unabsorbed region of the spectrum for $f_X=1$ (blue squares) and $f_X=0.1$ (black diamonds), plotted against the number of absorption lines added to the spectrum.  Error bars are derived from a bootstrap resampling of each spectrum.}
\label{f:stddevsplot}
\end{figure}

\begin{figure}
\begin{center}
\resizebox{\columnwidth}{!}{\includegraphics[angle=0]{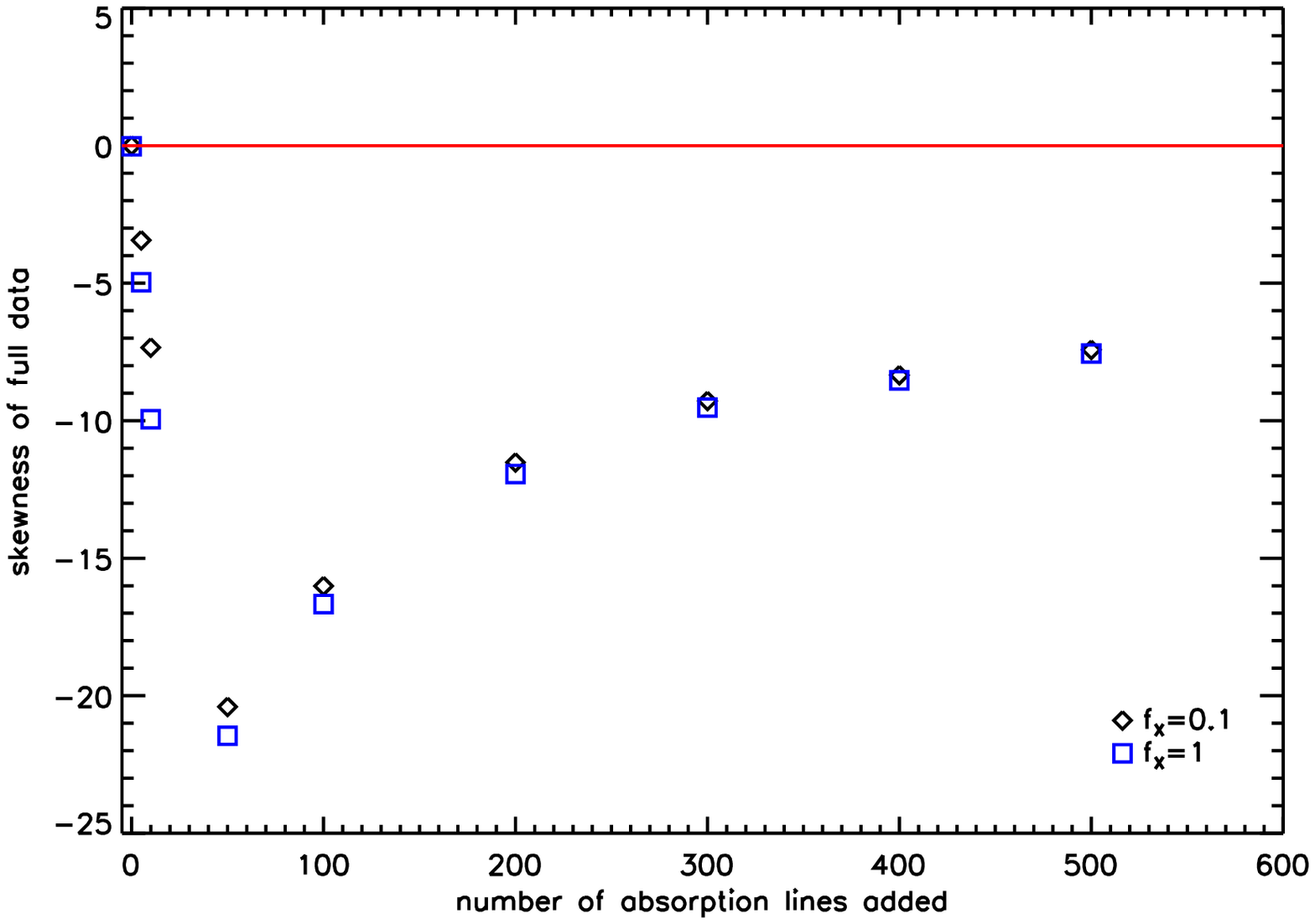}}
\end{center}
\caption{Skewness of the distribution of the normalized flux values in the absorbed region for $f_X=1$ (blue squares) and $f_X=0.1$ (black diamonds), plotted against the number of absorption lines added to the spectrum.}
\label{f:skewsplot}
\end{figure}

In Figures \ref{f:meansplot}, \ref{f:stddevsplot} and \ref{f:skewsplot}, we show the Gaussian fit-derived flux decrement, the difference between the absorbed and unabsorbed fit-derived standard deviation, and the simulation points' skewness as a function of the number of lines added.  In each plot, blue squares indicate $f_X=1$, black diamonds indicate $f_X=0.1$, and a red line is included at 0 for reference.  For Figures \ref{f:meansplot} and \ref{f:stddevsplot}, we estimated the errors of the Gaussian fit parameters using bootstrap resampling of the spectrum points.

Figure \ref{f:meansplot} shows that, in the idealized case in which foregrounds and the source spectra are well understood, this method can very accurately determine $f_X$.  The robustness of the flux decrement estimated by the Gaussian fit -- regardless of the number of lines included -- is striking.  We see in the plot that even for $f_X=1$, the flux decrement is easily distinguished from 0, making a detection of 21cm absorption and an estimate of $f_X$ possible.  Even if there are 500 lines added to the absorption spectrum, the flux decrement can be translated into a high precision estimate of $\tau$ via equation (\ref{eq:taufromfluxdec}).  For $f_X=0.1$, the flux decrement is similarly stable and is detectable at a much higher significance.  The strong separation between the $f_X=1$ and $f_X=0.1$ points in Figure \ref{f:meansplot} further suggests that $f_X$ might potentially be determinable to closer than an order of magnitude if the foregrounds and spectra were well known.

Figure \ref{f:stddevsplot} plots $\sigma_R - \sigma_L$ where $\sigma_R$ is the Gaussian-derived standard deviation in the absorbed region and $\sigma_L$ is the value in the unabsorbed region.  It is clear from this figure that while the outlying absorption feature points do not significantly move the Gaussian, they do somewhat broaden it (though not significantly enough that an $f_X=1$ spectrum with many lines would be mistaken for an $f_X=0.1$ spectrum with few lines). This measurement should be possible even in a realistic observation in which the foregrounds and source spectrum had to be approximated with a running powerlaw and extrapolated from the unabsorbed region, as the relative widths of the distributions should still be robust.

Figure \ref{f:skewsplot} presents the skewness of the simulation distribution as a function of lines added.  As can be seen in the plot, a large negative skewness occurs when large numbers of absorption lines are present.  The skewness is not monotonic, however.  This is due to the fact that the presence of more absorption lines means there are fewer channels that contribute to the roughly Gaussian part of the flux distribution, and although the distribution is more heavily skewed at the low-flux end, it also becomes broader and less Gaussian overall.  Since the standard deviation contributes inversely in the definition of skewness, the magnitude of the skewness decreases as the number of outliers grows very large.

% Histograms and fits for no lines and 100 lines added

% Gaussian means plot

% Gaussian standard deviations plot

% Skewnesses plot

\section{Conclusions and outlook}
\label{s:conclusions}

We have shown that the magnitude and statistical significance of the 21cm forest signal is highly sensitive to assumptions about the thermal history of the IGM, and that the detectability of the signal also depends strongly on the redshift and radio-loudness of the background source.  The strongest signal can be seen when x-ray heating is relatively inefficient (as parameterized by low values of the x-ray efficiency $f_X$, described in Sections \ref{s:numsims} and \ref{s:fx}) and when the source is very distant and radio-loud.  For example, if the IGM is heated in a process for which $f_X=1$, a background radio source with flux density $S \gtrsim 100$ mJy at redshift $z \gtrsim 10$ would be needed for 21cm forest features to be significantly detected within a week of integration on the SKA.  If x-ray heating is relatively inefficient, however, and $f_X \approx 0.01$, the same signal-to-noise could be achieved with a source of 10 mJy at redshift $z \approx 9$.  In any case, it is clear that radio-loud sources at high redshift must be found for the 21cm forest signal to be observed.

This conclusion is based on a relation [equation (\ref{eq:Smin})] commonly used to quantify the detectability of the 21cm forest.  However, we point out that the relation applies specifically to attempts to study the features of the 21cm forest at high signal-to-noise.  In this study, our main interest is in using the 21cm forest to study the thermal history of the high-redshift IGM.  We find this can be accomplished through the use of statistical methods that allow a high-significance detection of 21cm absorption against objects with lower flux densities or redshifts than those suggested by equation (\ref{eq:Smin}).  This suggests that the detection of 21cm absorption may be less challenging than previously thought.

One of the statistical methods we discuss, which is applicable when strong absorption lines are not present, makes use of the increase in variance in the source spectrum at the frequency at which 21cm absorption begins.  The passage of the radio signal alternately through regions of high ionization (HII regions produced by photoionization from bright UV sources) and through regions of relatively neutral gas (which can strongly absorb 21cm radiation) results in a pattern of high and low transmission and thus significantly increases the variance of the signal above the noise in certain cases.  The measurement of this effect may allow us to detect 21cm absorption and place limits on the thermal history of the IGM even in cases where the detection of the forest features is not significant.  A combination of long integration times and variance studies can therefore allow us to extract information from the 21cm forest even if the IGM is relatively hot and if bright high-redshift sources are rare.

In the case that strong absorption lines are numerous, the lines dominate the spectrum's variance in the absorbed region, making the use of the variance method less straightforward.  Although an increase in variance can signal that absorption is occurring \citep{Carilli:2004}, the uncertainties in modelling the number and character of absorption lines make it difficult to directly link the level of variance increase with the thermal properties of the high-redshift IGM.  A measurement of the average flux decrement in the absorbed region would similarly be strongly affected by the properties of the absorption lines.

In Section \ref{s:gauss}, we present a second method that can be used to extract thermal information about the IGM even if a spectrum contains many strong absorption lines.  The method effectively removes absorption lines based on their narrowness and non-Gaussian nature, by fitting the source-subtracted flux distribution across a wide band to a Gaussian. We show that the flux decrement determined with this method can be detected at high significance even for observational scenarios far less favorable than those derived through a naive application of equation (\ref{eq:Smin}).  
The estimates here should be considered a best-case scenario, due to the difficulty of accurately source-subtracting the spectrum in the presence of foregrounds.
Equation (\ref{eq:Smin}) suggests that an SKA-like observatory looking for absorption against a $z=9$ source with a radio flux similar to Cygnus A would require nearly a decade of integration time for a detection of the 21cm forest.  By contrast, our method shows that a significant detection of the flux decrement -- and therefore a measurement of the optical depth and a constraint on $f_X$ -- could, in principle, be achieved with an integration time of less than 1 week for the same object and observatory. Even without perfect source subtraction, the width of the flux distribution over a wide band can be used to clearly distinguish amongst different values of $f_X$.
Note that this does not depend strongly on properties of the absorption lines [in contrast to, e.g. \citet{Xu:2011}]. 

This has implications for near-term reionization-era observational efforts.  Our result shows that high-significance detections may be possible with instruments such as LOFAR or MWA, given a sufficiently radio-loud and distant source and a good understanding of foregrounds and source spectra.
Looked at a different way, this method could make it possible for the 21cm forest to be detected against lower-redshift or dimmer sources with the SKA.

In this work, we aimed to present the 21cm forest signal produced by the mean evolving IGM and the structure of large-scale photoionization along the line of sight.  
While this study presents an outline of the feasibility and major challenges of future 21cm forest measurements of the IGM, another potential extension of this work would be to use a more detailed simulation of the IGM properties and intervening absorption lines, perhaps from a large-scale numerical simulation including realistic dark matter and baryonic physics and radiative transfer, to more precisely simulate the expected signal in a high-redshift radio source.  Planned future work also involves a more detailed treatment of the scope of this method in light of the capability of near-term observatories, and the possibility of fitting for the reionization model as part of the analysis.  Another crucial element of any estimate of the utility of 21cm forest studies is an accurate estimate of the number of high-redshift radio-loud sources that will be available in the era of the SKA.  We have given a brief overview in Section \ref{s:sourcepop} of recent estimates, but updated theoretical studies and more observational follow-up of known radio-loud sources would both strongly impact the case for pursuing 21cm forest observations as a diagnostic tool for studying the reionization-era IGM.

\section{Acknowledgments}

Simulations used in this work were run by Paul Geil, Sudhir Raskutti and Lila Warszawski.  KJM would like to thank James Bolton, Sebastiano Cantalupo, Andrew Pontzen, and Debora Sijacki for useful discussions.  KJM also thanks the Melbourne University School of Physics Astro Group for their hospitality while the bulk of this work was carried out.  We thank the anonymous referee for comments which have improved this work.

\label{lastpage}


\begin{thebibliography}{}

\bibitem[Bolton \& Haehnelt(2007)]{Bolton:2007}
Bolton, J.S. \& Haehnelt, M.G. 2007, MNRAS 382, 325.

\bibitem[Carilli et al.(2004)]{Carilli:2004}
Carilli, C.L. 2004, New Astron. Rev. 48, 1053.

\bibitem[Carilli, Gnedin \& Owen(2002)]{Carilli:2002}
Carilli, C.L., Gnedin, N.Y. \& Owen, F. 2002, ApJ 577, 22.

\bibitem[Carilli(2006)]{Carilli:2006}
Carilli, C.L. 2006, NAR 50, 162.

\bibitem[Furlanetto(2006a)]{Furlanetto:2006a}
Furlanetto, S.R. 2006a, MNRAS 370, 1867.

\bibitem[Furlanetto(2006b)]{Furlanetto:2006b}
Furlanetto, S.R. 2006b, MNRAS 371, 867.

\bibitem[Furlanetto \& Loeb(2002)]{Furlanetto:2002}
Furlanetto, S.R. \& Loeb, A. 2002, ApJ 579, 1.

\bibitem[Furlanetto, Oh \& Briggs(2006)]{FOB}
Furlanetto, S.R., Oh, S.P. \& Briggs, F.H. 2006, Phys. Rep. 433, 181.

\bibitem[Geil \& Wyithe(2008)]{Geil:2008}
Geil, P.M. \& Wyithe, J.S.B. 2008, MNRAS 386, 1683.

\bibitem[Haiman et al.(2004)]{Haiman:2004}
Haiman, Z., Quataert, E. \& Bower, G. 2004, ApJ 612, 698.

\bibitem[Ivezic et al.(2002)]{Ivezic:2002}
Ivezic, Z. et al. 2001, ApJ 124, 2364.

\bibitem[Jiang et al.(2007)]{Jiang:2007}
Jiang et al. 2007, ApJ 656, 680.

\bibitem[Larson et al.(2010)]{Larson:2010}
Larson, D. et al. arxiv:1001.4635

\bibitem[Mack \& Wesley(2008)]{MackWesley}
Mack, K.J. \& Wesley, D.H., arxiv:0805.1531

\bibitem[McQuinn et al.(2007)]{McQuinn:2007}
McQuinn, M. et al. 2007, MNRAS 377, 1043.

\bibitem[Mitra, Choudhury \& Ferrara(2010)]{Mitra:2010}
Mitra, S., Choudhury, T.R. \& Ferrara, A. arxiv:1011:2213

\bibitem[Parsons et al.(2010)]{Parsons:2010}
Parsons, A.R. et al. 2010, AJ 139, 1468.

\bibitem[Pritchard \& Loeb(2008)]{Pritchard:2008}
Pritchard, J. \& Loeb, A. 2008, Phys. Rev. D 78, 103511.

\bibitem[Pritchard \& Loeb(2010)]{Pritchard:2010}
Pritchard, J. \& Loeb, A. 2010, Phys. Rev. D 82, 023006.

\bibitem[Santos et al.(2008)]{Santos:2008}
Santos, M.G. et al. 2008, ApJ 689, 1.

\bibitem[Xu et al.(2009)]{Xu:2009}
Xu, Y. et al. 2009, ApJ 704, 1396.

\bibitem[Xu, Ferrara \& Chen(2011)]{Xu:2011}
Xu, Y., Ferrara, A. \& Chen, X. 2011, MNRAS 410, 2025.

\end{thebibliography}
\end{document}